\documentclass[a4j,11pt]{article}

\usepackage{amsmath}
\usepackage{amsmath,amssymb,bm}
\usepackage[dvipdfm]{graphicx, color}

\title{Heavy gravitino, naturalness, and sizable anomaly mediation}

\author{
\centerline{
Nobuhiro~Maekawa$^{1,2}$\footnote{E-mail address: maekawa@eken.phys.nagoya-u.ac.jp}
~and 
Kenichi~Takayama$^{1}$\footnote{E-mail address: takayama@th.phys.nagoya-u.ac.jp}}
\\*[25pt]
\centerline{
\begin{minipage}{\linewidth}
\begin{center}
$^1${\it \normalsize Department of Physics, Nagoya University, Nagoya 464-8602, Japan }  \\*[10pt]
$^2${\it \normalsize Kobayashi Maskawa Institute, Nagoya University, Nagoya 464-8602, Japan }  \\*[10pt]\end{center}
\end{minipage}}
\\*[50pt]}

\date{}

\begin{document}

\maketitle

\begin{abstract}

We consider the situation in which $m_{3/2}\sim O(100{\rm TeV})$ for solving the gravitino problem and the other supersymmetry(SUSY) breaking parameters are $O(1{\rm TeV})$ for the naturalness.
We point out that the anomaly mediation cancels out the renormalization group contribution to the gaugino masses and the sfermion masses other than the stop masses at a scale which is called the mirage scale. 
The situation is similar to the mirage mediation, in which special boundary conditions for the SUSY breaking parameters are required, though for the stop masses and the up-type Higgs mass, such cancellation at the mirage scale does not happen.
Despite no cancellation for the up-type Higgs mass, we show that the little hierarchy problem becomes less severe in this situation.
One advantage of this situation over the mirage mediation is that the stop mixing parameter $A_t$ can be larger and therefore, smaller stop mass is sufficient for 125 GeV Higgs.
When the mirage scale is around TeV scale, the SUSY breaking parameters induced by the gravity mediation at the grand unification scale can be observed directly by the TeV scale experiments.

\end{abstract}

\vspace{1cm}

\section{Introduction}

The minimal supersymmetric (SUSY) Standard Model (MSSM) is still one of the most promizing 
candiates as physics beyond the Standard model (SM). 
The MSSM can solve the gauge hierarchy problem and provide a dark matter candidate as the lightest
 supersymmetric particle (LSP).
Moreover, the SUSY grand unified theory (GUT) is experimentally supported by
the remarkable coincidence of three SM gauge coupling constants around $10^{16}$ GeV.
However, many SUSY models suffer from a tuning problem, called the SUSY little hierarchy problem.
This problem arises from a tension between naturalness which requires lightness of several SUSY
 particles and the Higgs mass $m_h=125$ GeV\cite{Aad:2012tfa, Chatrchyan:2012ufa} which forces those to be heavy.
Cosmologically, it has been pointed out that the decay of the gravitino spoils the success of the Big Bang Nucleosynthesis (BBN). This is called the gravitino problem\cite{gravitino, gravitino2}. 

One of the simplest solution for the gravitino problem is to assume that the gravitino decays before the BBN begins. For example, if the gravitino is heavier than 100 TeV, then the lifetime of the gravitino becomes of order $10^{-2}$ sec. At the time in the history of the universe, the proton-neutron ratio has not been fixed yet by freezing out the weak decay process.
In the literature, the high scale SUSY breaking scenario\cite{HighScale}, in which the scaler fermion masses are taken to be the same order of the heavy gravitino mass, has been studied because such scenario can realize the Higgs mass $m_h\sim 125$ GeV without the large stop mixing parameter $A_t$\cite{HighScale2}. 
Such high scale SUSY breaking scenario has various advantages, for example, it has no SUSY flavor problem, no SUSY CP problem, etc. 
Unfortunately the fine-tuning problem on the Higgs mass becomes much worse in the scenario.

For the fine-tuning problem, it is preferable that the stop masses and the gaugino masses are of order 1 TeV.
These two requirements, the gravitino mass $m_{3/2}\geq 100{\rm TeV}$ and the sfermion masses $\tilde m \sim O(1{\rm TeV})$, are not inconsistent with each other. Actually, both requirements are satisfied in the mirage mediation scenario\cite{mirage, mirage2} in which the moduli\cite{moduli} and anomaly\cite{anomaly} contributions to SUSY breaking parameters become comparable.
One of the most important features in the mirage mediation is that the effective SUSY mediation scale can be lower because the renormalization group effects can be cancelled by the effect of the anomaly mediation.
As a result, the little hierarchy problem may be solved\cite{littlemirage, littlemirage2}. 
Unfortunately, in the mirage mediation, very specific boundary conditions for the SUSY breaking parameters are required.
What happens if we take more generic boundary conditions for the SUSY breaking parameters?
If the contribution of the anomaly mediation dominates that of the gravity mediation, then the mass squares of the right-handed slepton become negative. Therefore, we have an upper bound for the 
gravitino mass, which is nothing but $O(100{\rm TeV})$. 

In this paper, we will examine a scenario in which the gravitino mass is of order 100 TeV to solve the gravitino problem and the other SUSY breaking parameters, which are induced by the gravity mediation, are around the TeV scale to stabilize the weak scale. 
We will not discuss how to realize such a situation.
Here we simply note that it can be possible at least in mirage 
mediation scenario.

Let us examine the little hierarchy problem in more detail, because it is one of the main purposes in this paper to improve the fine-tuning in the Higgs sector.
In supersymmetric models, a quantum correction of the up-type Higgs squared mass $\Delta m^2_{H_u}$ strongly depends on the stop mass $m_{\tilde{t}}$\ : 
\begin{equation}
\Delta m^2_{H_u}\sim -\frac{3y_t^2}{8\pi^2}(m^2_{\tilde{t}_L}+m^2_{\tilde{t}_R}+A_t^2)\ln\frac{\Lambda}{m_{\tilde{t}}},
\end{equation}
where $\Lambda$ is the messenger scale and here we consider $\Lambda=2\times 10^{16}$ GeV.
In order to realize the electroweak symmetry breaking without the fine-tuning, one can expect that $m_{\tilde{t}}$ is order of 100 GeV.
On the other hand, the lightest CP-even Higgs mass $m_h$ is also linked to the stop mass: 
\begin{equation}
m^2_h\simeq m_Z^2\cos^2 2\beta + \frac{3G_Fm_t^4}{\sqrt{2}\pi^2}\left[\ln\frac{m^2_{\tilde{t}}}{m^2_t}+\frac{A_t^2}{m^2_{\tilde{t}}}(1-\frac{A_t^2}{12m^2_{\tilde{t}}})\right].
\end{equation}
The Higgs mass $m_h=125$ GeV, which is discovered by ATLAS and CMS, implies heavy stop mass such as several TeV.
Therefore, it is difficult to get the realistic Higgs mass without destroying naturalness.

One of the solutions to avoid the little hierarchy problem is to move beyond the MSSM.
For instance, one may add an extra singlet as in the next-to MSSM\cite{NMSSM}.
On the other hand, we can also reduce fine-tuning within the MSSM by lowering the messenger scale $\Lambda$ such as the low-scale gauge mediation model\cite{gauge}.
One can also lower the messenger scale effectively in the case where several SUSY breaking contributions cancel the renormalization group (RG) evolution as in the TeV-scale mirage mediation model
\cite{mirage, littlemirage, littlemirage2}.
Note that the large stop mass spoils the naturalness even if the messenger scale is small.
The value of the $m_{\tilde{t}}$ with realizing the 125 GeV Higgs depends on the value of the $A_t$.
It is minimized when $|A_t/m_{\tilde{t}}|=\sqrt{6}$\cite{maximal}. 
It is, however, difficult to realize the large $A_t$ in the low-scale messenger models.
In the TeV-scale mirage mediation, the model fixes the ratio $A_t/m_{\tilde{t}}=\sqrt{2}$ at the mirage scale, which is considered to be around the TeV scale.
The gauge mediation model also fails to get the large $A_t$ because it does not appear at the leading order.
The value of $A_t$ in these models is not sufficient to get the Higgs mass naturally.

What happens if we do not impose the specific condition $A_t/m_{\tilde{t}}=\sqrt{2}$ in the mirage mediation?
To answer this question, we have to know what happens when the specific boundary conditions in the mirage mediation scenario are not imposed. 
This is one of our motivation for the work in this paper.

The paper proceeds as follows.
In section 2, we recall that the anomaly mediation contribution
can cancel the RG evolution of the gravity mediation contribution by the analytic solutions of
one-loop RG equations of the MSSM. In section 3, we study what happens if the gravity mediation
produces $O(1 {\rm TeV})$ SUSY breaking parameters while the gravitino mass is $O(100 {\rm TeV})$.
Especially, we show that the little hierarchy problem becomes less severe like in the mirage 
mediation. And section 4 is for the discussion and summary.

\section{Cancellation property of the anomaly mediation}

It is known that the anomaly mediation\cite{anomaly} has the property to cancel the RG evolution of the gravity mediation. In this section, we will review this property by solving the one-loop RG equations for the SUSY breaking parameters in the MSSM.

\subsection{Small Yukawa case}

Let us see this cancellation property in the case where the Yukawa coupling can be neglected.
The results in this subsection can be applied to all sfermion masses except stop masses and up-type Higgs mass $m_{H_u}$ when bottom and tau Yukawa couplings can be neglected, i.e., $\tan\beta\equiv \langle H_u\rangle /\langle H_d\rangle\ll 50$.

First we consider the gaugino mass $M_a$ $(a=1,2,3)$. 
It satisfies the RG equation
\begin{equation}
\frac{d}{dt}M_a=\frac{1}{8\pi^2}b_ag_a^2M_a
\label{eq:gaugino_RGeq}
\end{equation}
at one-loop level.
Here the gauge coupling $g_a$ obeys the RG equation
\begin{equation}
\frac{d}{dt}g_a=\frac{1}{16\pi^2}b_ag_a^3,
\end{equation}
where $(b_1, b_2, b_3)=(\frac{33}{5}, 1, -3)$ in the MSSM. 
Then the anomaly mediation solution is written as
\begin{equation}
M_a(\mu)|_{\mathrm{anomaly}}=\frac{1}{16\pi^2}b_ag_a^2m_{3/2},
\end{equation}
where $m_{3/2}$ is the gravitino mass.
There is also the gravity mediation solution as follows:
\begin{equation}
M_a(\mu)|_{\mathrm{gravity}}=\tilde{M_a}+\frac{1}{8\pi^2}b_ag_a^2\tilde{M_a}\ln\frac{\mu}{\Lambda},
\label{eq:gaugino_gravity}
\end{equation}
where $\tilde{M_a}$ is the mass from the gravity mediation at the cutoff scale $\Lambda$.
Note that, in this paper,  ``the gravity mediation'' does not include the anomaly mediation.
Hereafter we assume that $\tilde{M}_a$ is universal as
\begin{equation}
\tilde{M}_1=\tilde{M}_2=\tilde{M}_3=M_{1/2},
\end{equation}
which is imposed if the GUT is assumed at the cutoff scale 
$\Lambda=\Lambda_G=2\times 10^{16}$ GeV.
One can easily check that these two expressions satisfy the RG equation (\ref{eq:gaugino_RGeq}), respectively.
These two contributions can coexist because the sum $M_a|_\mathrm{anomaly}+M_a|_\mathrm{gravity}$ also satisfies the same RG equation.
It can be rewritten as
\begin{equation}
M_a(\mu)=M_{1/2}+\frac{1}{8\pi^2}b_ag_a^2M_{1/2}\ln\frac{\mu}{M_{\mathrm{mir}}},
\label{eq:gaugino_total}
\end{equation}
where the mirage scale $M_{\mathrm{mir}}$ is defined as
\begin{equation}
\ln\frac{M_{\mathrm{mir}}}{\Lambda}=-\frac{m_{3/2}}{2M_{1/2}	}.
\end{equation}
At the mirage scale the anomaly mediation contribution cancel the quantum corrections of the gravity mediation contribution and we get $M_a(M_{\mathrm{mir}})=M_{1/2}$.

Second we see the trilinear coupling $A_{ijk}$.
The one-loop RG equation is
\begin{equation}
\frac{d}{dt}A_{ijk}=-\frac{1}{4\pi^2}\sum_a(C_i^a+C_j^a+C_k^a)g_a^2M_a,
\end{equation}
where $C_i^a$ is the quadratic Casimir coefficient for the field $i$ and $C_i^a=(N^2-1)/(2N)$ for a fundamental representation of the gauge group $SU(N)$, $C_i^a=q_i^2$ for the $U(1)$ charge $q_i$.
It is related to the anomalous dimension $\gamma_i$ as $\gamma_i=2\sum_a C_i^ag_a^2$.
Then the anomaly mediation
\begin{equation}
A_{ijk}(\mu)|_\mathrm{anomaly}=-\frac{1}{16\pi^2}(\gamma_i+\gamma_j+\gamma_k)m_{3/2}
\end{equation}
and the gravity mediation
\begin{equation}
A_{ijk}(\mu)|_\mathrm{gravity}=\tilde{A}_{ijk}-\frac{1}{8\pi^2}(\gamma_i+\gamma_j+\gamma_k)M_{1/2}\ln\frac{\mu}{\Lambda}
\end{equation}
satisfiy the RG equation when they are combined with the $M_a|_\mathrm{anomaly}$ and $M_a|_\mathrm{gravity}$, respectively.
Here $\tilde{A}_{ijk}$ are also the gravity mediation contribution at the cutoff scale.
The sum of two contributions $(A_{ijk}|_\mathrm{anomaly}+A_{ijk}|_\mathrm{gravity}, M_a|_\mathrm{anomaly}+M_a|_\mathrm{gravity})$ also obeys the same RG equation.
As a result, 
\begin{equation}
A_{ijk}(\mu)=\tilde{A}_{ijk}-\frac{1}{8\pi^2}(\gamma_i+\gamma_j+\gamma_k)M_{1/2}\ln\frac{\mu}{M_{\mathrm{mir}}}.
\end{equation}
One can see that the RG evolution of the trilinear coupling also vanishes at $M_{\mathrm{mir}}$.

Lastly we see the scalar mass $m_i^2$.
The one-loop RG equation is
\begin{equation}
\frac{d}{dt}m_i^2=-\frac{1}{2\pi^2}\sum_aC_i^ag_a^2|M_a|^2+\frac{3}{40\pi^2}g_1^2Y_iS,
\end{equation}
where the quantity $S$ is defined as
\begin{equation}
S=\sum_i Y_im_i^2=m^2_{H_u}-m^2_{H_d}+\mathrm{Tr}\,[m^2_{\tilde{Q}}-2m^2_{\tilde{u}_R}+m^2_{\tilde{d}_R}-m^2_{\tilde{L}}+m^2_{\tilde{e}_R}].
\end{equation}
The scalar mass is generated from the anomaly mediation and the gravity mediation as
\begin{equation}
m^2_i(\mu)|_{\mathrm{anomaly}}=-\frac{1}{32\pi^2}\dot{\gamma}_im^2_{3/2}
\end{equation}
\begin{equation}
m^2_i(\mu)|_{\mathrm{gravity}}=\tilde{m}^2_i-\frac{1}{4\pi^2}\gamma_iM_{1/2}^2\ln\frac{\mu}{\Lambda}-\frac{1}{8\pi^2}\dot{\gamma}_iM_{1/2}^2\left(\ln\frac{\mu}{\Lambda}\right)^2+\frac{3}{40\pi^2}Y_ig_1^2\tilde{S}\ln\frac{\mu}{\Lambda},
\label{eq:scalar_gravity}
\end{equation}
where $\dot{\gamma}_{i}=\frac{d}{dt}\gamma_i$, $\tilde{S}=\sum_iY_i\tilde{m}^2_i$ and  $\tilde{m}^2_i$ is the mass from the gravity mediation at the cutoff scale.
They satisfy the RG equation when they are combined with the $M_a|_\mathrm{anomaly}$ and $M_a|_\mathrm{gravity}$, respectively.
However, the combination $(m_i^2|_\mathrm{anomaly}+m_i^2|_\mathrm{gravity}, M_a|_\mathrm{anomaly}+M_a|_\mathrm{gravity})$ does not satisfy the same RG equation.
It is not the problem because the scalar mass has interference terms
\begin{equation}
m^2_i(\mu)|_{\mathrm{interference}}=-\frac{1}{8\pi^2}\gamma_iM_{1/2}m_{3/2}-\frac{1}{8\pi^2}\dot{\gamma}_iM_{1/2}m_{3/2}\ln\frac{\mu}{\Lambda}
\end{equation}
when there are the different SUSY breaking sources.
It guarantees the coexistence of the two contributions.
After all, the scalar mass under the anomaly mediation and the gravity mediation is
\begin{equation}
m^2_i(\mu)=\tilde{m}^2_i-\frac{1}{4\pi^2}\gamma_iM_{1/2}^2\ln\frac{\mu}{M_{\mathrm{mir}}}-\frac{1}{8\pi^2}\dot{\gamma}_iM_{1/2}^2\left(\ln\frac{\mu}{M_\mathrm{mir}}\right)^2+\frac{3}{40\pi^2}Y_ig_1^2\tilde{S}\ln\frac{\mu}{\Lambda}.
\label{eq:scalar_total}
\end{equation}
Note that the RG evolution of the scalar mass also cancels at $M_\mathrm{mir}$ if $\tilde{S}$ vanishes.
Hereafter we assume $\tilde{S}=0$ because it is satisfied in the GUT models where $H_u$ and $H_d$ are unified into a single multiplet, such as $SO(10)$.

We have seen that all the RG evolution effects of gaugino mass, trilinear coupling and scalar mass vanish at the same scale $M_{\mathrm{mir}}$ in the small Yukawa case.
Therefore we can see that the anomaly mediation effectively lowers the cutoff scale $\Lambda$ to $M_{\mathrm{mir}}$.
In the case of $m_{3/2}/M_{1/2}\sim 60$, the mirage scale is around the TeV scale.
Note that the value $m_{3/2}/M_{1/2}\sim 60$ is consistent with the assumptions, $m_{3/2}\sim 100$ TeV, which is for solving the gravitino problem and that the SUSY breaking scale is around 1 TeV.

\subsection{Effect of top Yukawa}
	
We have showed that the anomaly mediation cancels the RG evolution of the gravity mediation if there is no Yukawa coupling in the previous subsection.
However, the expressions for $m^2_{H_u}$, $m^2_{\tilde{t}_L}$, $m^2_{\tilde{t}_R}$ and $A_t$ should be modified because the top Yukawa coupling has the sizable contribution.
Here we consider the case where the bottom and tau Yukawa coupling contributions can be neglected.

Let us see the effect of the top Yukawa coupling in more detail.
First, the RG equation of the top Yukawa coupling is
\begin{equation}
\frac{d}{dt}y_t=\frac{1}{16\pi^2}y_t(6y_t^2-2\sum_aC_t^ag_a^2)
\end{equation}
with $C_t^a=C^a_{t_L}+C^a_{t_R}+C^a_{H_u}$.
The running top Yukawa coupling is given as
\begin{equation}
y_t^2(\mu)=\frac{y_t^2(\Lambda)E(\mu)}{1-\frac{3}{4\pi^2}y_t^2(\Lambda)F(\mu)},
\end{equation}
where the function $E(\mu)$ and $F(\mu)$ are defined as
\begin{equation}
E(\mu)=\prod_a\left(1-\frac{b_a}{8\pi^2}g_{\mathrm{GUT}}^2\ln\frac{\mu}{\Lambda}\right)^{2C_t^a/b_a}
\end{equation}
\begin{equation}
F(\mu)=\int_\Lambda^\mu\frac{d\mu'}{\mu'}E(\mu').
\end{equation}
The RG equations for the $A_t$ and $m^2_i$ $(i=\tilde{t}_L, \tilde{t}_R, H_u)$ become
\begin{equation}
\frac{d}{dt}A_t=\frac{1}{4\pi^2}\left(3|y_t|^2A_t-\sum_aC_t^ag_a^2M_a\right)
\end{equation}
\begin{equation}
\frac{d}{dt}m_i^2=-\frac{1}{8\pi^2}\left(k_i|y_t|^2(m^2_{H_u}+m^2_{\tilde{t}_L}+m^2_{\tilde{t}_R})+k_i|y_t|^2|A_t|^2+4\sum_aC_i^ag_a^2|M_a|^2\right).
\end{equation}

With the top Yukawa coupling, the $A_t$, up-type Higgs mass and the stop masses generated by the gravity mediation are given as
\begin{equation}
A_{t}(\mu)=\tilde{A}_t+6\rho(\tilde{A}_t-M_{1/2})-\frac{1}{8\pi^2}(\gamma_{H_u}+\gamma_{t_L}+\gamma_{t_R})M_{1/2}\ln\frac{\mu}{\Lambda},
\label{eq:At_gravity}
\end{equation}
\[
m^2_i(\mu)=\tilde{m}^2_i-k_i\rho\left[(\tilde{A}_t-M_{1/2})^2(1+6\rho)+\tilde{\Sigma}_t-M_{1/2}^2\right]
\]
\begin{equation}
-\frac{M_{1/2}}{4\pi^2}\left[\gamma_iM_{1/2}+k_i(\tilde{A}_t-M_{1/2})(1+6\rho)y_t^2\right]\ln\frac{\mu}{\Lambda}-\frac{1}{8\pi^2}\dot{\gamma}_iM_{1/2}^2\left(\ln\frac{\mu}{\Lambda}\right)^2,
\label{eq:mi_gravity}
\end{equation}
where $\tilde{\Sigma}_t=\tilde{m}^2_{H_u}+\tilde{m}^2_{\tilde{t}_L}+\tilde{m}^2_{\tilde{t}_R}$.
The anomalous dimension $\gamma_i$ is written as
\begin{equation}
\gamma_i=2\sum_aC_i^a g_a^2+k_iy_t^2,
\end{equation}
where $k_{H_u}=-3, k_{t_L}=-1, k_{t_R}=-2$ and $k_i=0$ for the other fields.
The effect of the top Yukawa coupling is involved in the parameter $\rho$:
\begin{equation}
\rho(\mu)=\frac{y_t^2(\mu)}{8\pi^2}\frac{F(\mu)}{E(\mu)}.
\end{equation}
Note that if the function $E(\mu)$ is just a constant, $\rho$ can be estimated as $\rho\sim \ln(\mu/\Lambda)$.

The anomaly mediation changes the expressions (\ref{eq:At_gravity}) and (\ref{eq:mi_gravity}) as
\begin{equation}
A_{t}(\mu)=\tilde{A}_t+6\rho(\tilde{A}_t-M_{1/2})-\frac{1}{8\pi^2}(\gamma_{H_u}+\gamma_{t_L}+\gamma_{t_R})M_{1/2}\ln\frac{\mu}{M_{\mathrm{mir}}},
\end{equation}
\[
m^2_i(\mu)=\tilde{m}^2_i-k_i\rho\left[(\tilde{A}_t-M_{1/2})^2(1+6\rho)+\tilde{\Sigma}_t-M_{1/2}^2\right]
\]
\begin{equation}
-\frac{M_{1/2}}{4\pi^2}\left[\gamma_iM_{1/2}+k_i(\tilde{A}_t-M_{1/2})(1+6\rho)y_t^2\right]\ln\frac{\mu}{M_{\mathrm{mir}}}-\frac{1}{8\pi^2}\dot{\gamma}_iM_{1/2}^2\left(\ln\frac{\mu}{M_\mathrm{mir}}\right)^2.
\label{eq:mi_anomaly}
\end{equation}
These analytic formula are given by Ref.\cite{mirage2}.
One can see that the cancellation at the mirage scale is spoiled by the top Yukawa contribution.
Moreover, the large logarithmic factor appears because $\rho\sim \ln(\mu/\Lambda)$.
However, if we impose the special boundary condtions, $\tilde A_t=M_{1/2}=\sqrt{\tilde\Sigma_t}$, as in the mirage mediation scenario, then the cancellation at the mirage scale can be restored.
In the literature\cite{littlemirage, littlemirage2}, the improvement for the tuning has been discussed in the mirage mediation if the mirage scale is around the weak scale. 

What happens in the more general case in which the special boundary conditions are not satisfied?
We will discuss this subject in the next section.

\section{More general cases}

In the usual mirage mediation, the special boundary conditions for the gravity contribution to the SUSY breaking parameters are imposed, i.e., the universal sfermion masses to satisfy the flavor changing neutral current (FCNC) constraints, vanishing Higgs masses, and $\tilde A_t=M_{1/2}=\sqrt{\tilde\Sigma_t}$.
In this section, we study more general cases in which the anomaly mediation contribution is sizable.

\subsection{Generalization of mirage mediation: Natural SUSY}

Before going completely general cases, we discuss the cases in which the cancellation is complete as in the mirage mediation scenario.
In these cases, the little hierarchy problem can be quite improved as discussed in the usual mirage mediation.
It is obvious that for the cancellation, only the conditions $\tilde A_t=M_{1/2}=\sqrt{\tilde\Sigma_t}$ are important. For the cancellation, basically no additional condition is required for the other sfermion masses except vanishing $\tilde S$.

As an example, we mension the natural SUSY type boundary conditions\cite{NaturalSUSY}, in which the 
sfermion masses $m_3$ for the third generation ${\bf 10}$ of $SU(5)$ are around the TeV scale to stabilize the weak scale, and the other sfermion masses $m_0$ are taken to be much larger than $m_3$ to suppress the SUSY contributions to the FCNC processes and CP violation processes.
These boundary conditions are consistent with $\tilde A_t=M_{1/2}=\sqrt{\tilde\Sigma_t}$ and $\tilde S$ can vanish.
For example, we can adopt the  conditions $\tilde m_{H_u}^2=\tilde m_{H_d}^2=0$ and $\tilde m_{\tilde{Q}_3}^2=\tilde m_{\tilde{t}_{R}}^2=\tilde m_{\tilde{\tau}_R}^2=\tilde\Sigma_t/2$.
Similar boundary conditions in the mirage mediation, in which only stop masses are taken to be different from the others, have been discussed in the literature\cite{Naturalmirage}, though $\tilde S=0$ is not satisfied in their boundary conditions.
We think this possibility interesting because the $E_6$ GUT with the family symmetry $SU(2)_F$ predicts such natural SUSY type sfermion masses\cite{E6SU2}. 

The most important point is that if the mirage scale is around the SUSY breaking scale, we may directly obtain the signatures of GUT scenarios by observing the sfermion mass spectrum.
For example, if the rank of the unification group is larger than the rank of the SM gauge groups, the $D$-term contribution is non-vanishing generically. We may observe the magnitude of the $D$-term contribution directly.
In the usual arguments, by calculating the RG equations from the SUSY breaking scale to the GUT scale, we can obtain the signatures for the GUT scenarios from the observed sfermion mass spectrum\cite{KMY}. 
But in our cases, we do not have to calculate the RG equations or it is sufficient to calculate the RG flow a bit even if we have to. We will return to this point later.

\subsection{Upper bound for $m_{3/2}$ from stability conditions}

First of all, we explain the gravitino mass range which we would like to study in our scenario.
The lower bound of the gravitino mass is about 50 TeV\cite{gravitino2}, to solve the gravitino problem.
Strictly speaking, the lower bound is dependent on the reheating temperature of the inflation.
If low reheating temperature is considered, lower $m_{3/2}$ becomes possible. But if thermal leptogenesis is adopted for the baryogenesis, the lower bound of $m_{3/2}$ is not so different from 50 TeV.

\begin{figure}[tbp]
\begin{center}
\includegraphics[scale=0.26]{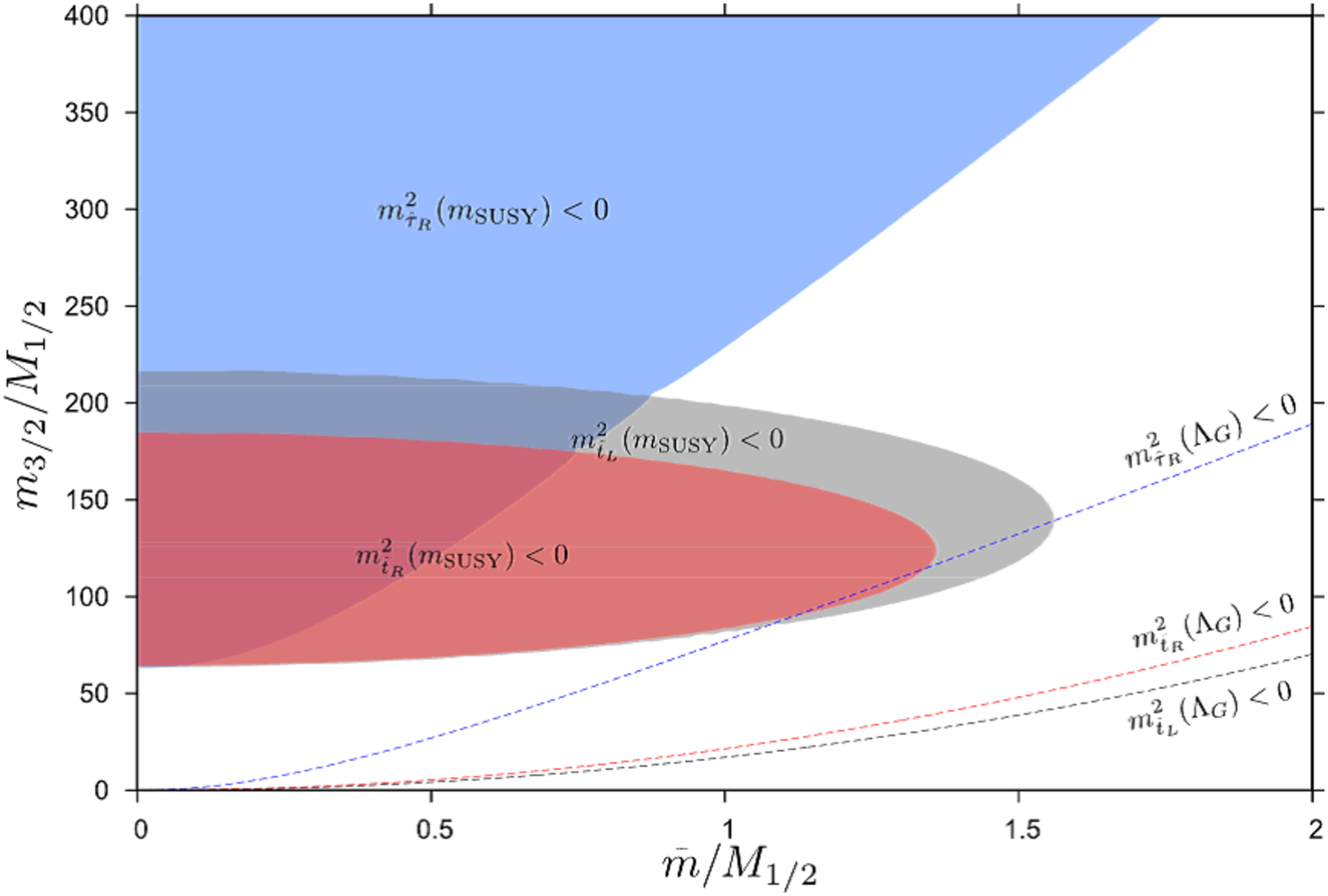}
\vspace{5mm}
\includegraphics[scale=0.26]{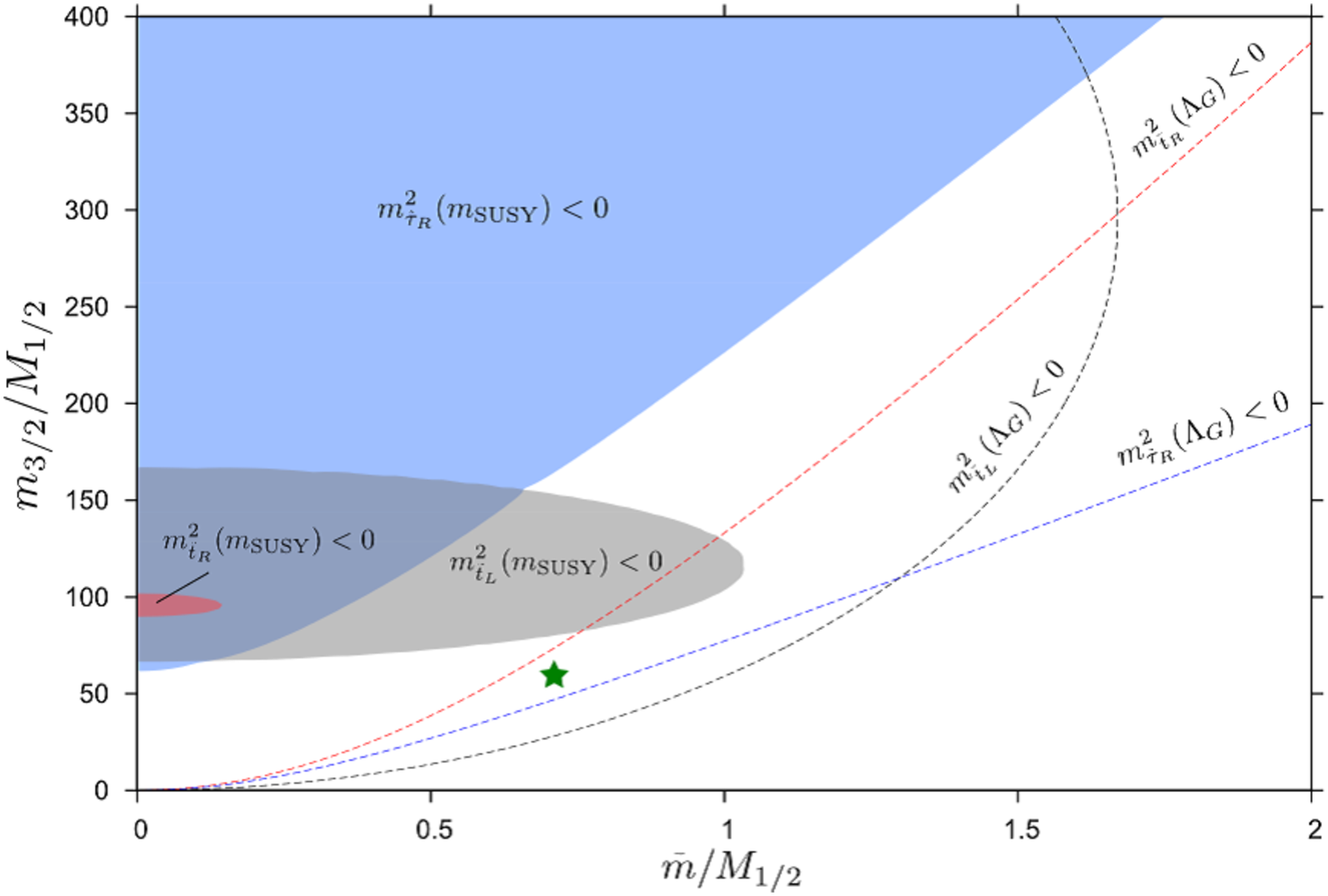}\
\vspace{5mm}
\includegraphics[scale=0.26]{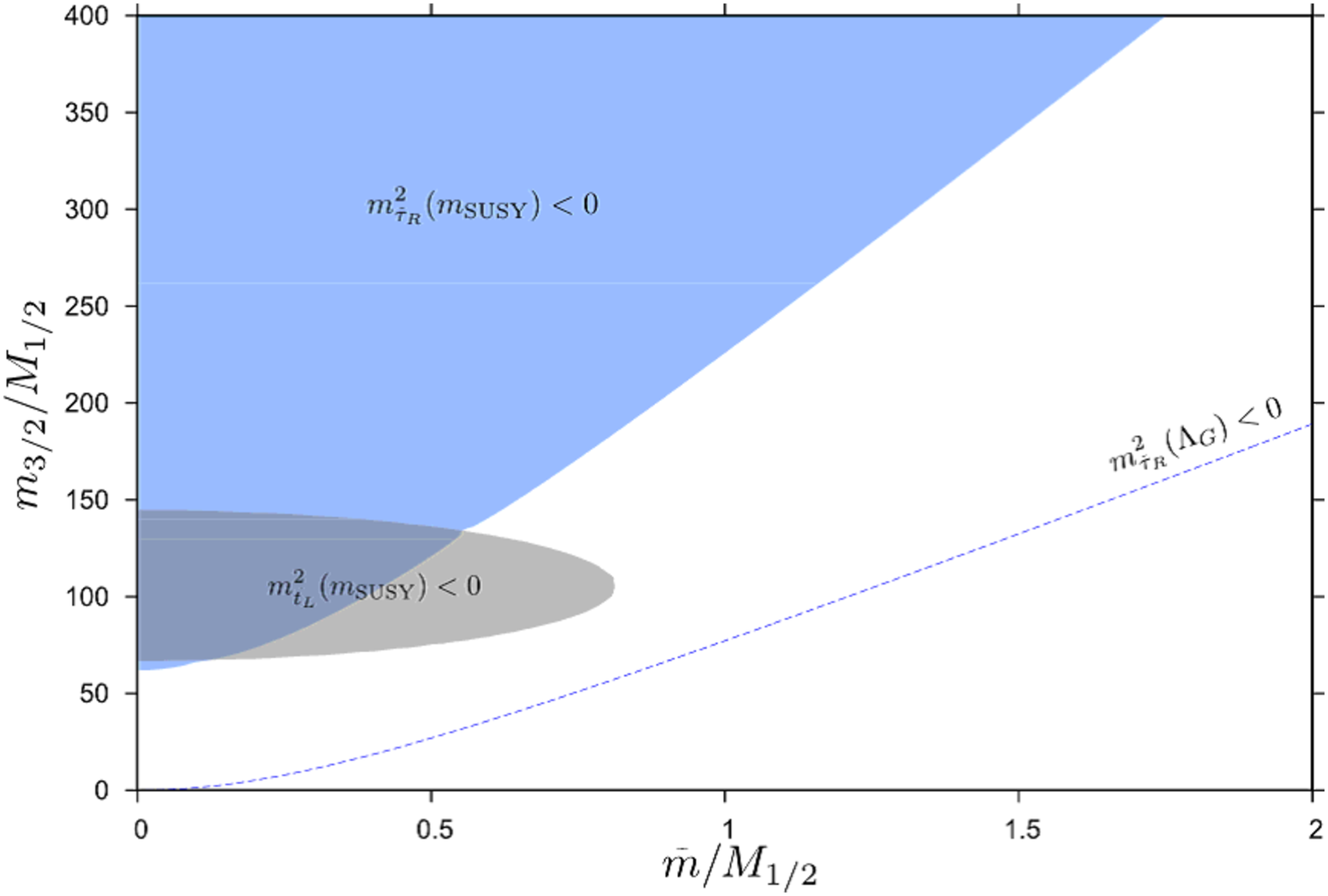}
\caption{Allowed region for the stability conditions at 1 TeV and at the GUT scale $\Lambda_G$ in
$(\tilde m/M_{1/2}, m_{3/2}/M_{1/2})$ plain, where 
$\tilde m=\tilde m_{\tilde t_L}=\tilde m_{\tilde t_R}=\tilde m_{\tilde \tau_R}$. 
The shaded region is forbidden by the stability conditions
at 1 TeV, and the upper side of the dotted line is the region where the mass square is negative at the
GUT scale. The upper left figure is for $\tilde A_t/M_{1/2}=-1$, the 
upper right figure is for $\tilde A_t/M_{1/2}=1$ and the lower figure is for $\tilde A_t/M_{1/2}=2$. The mirage point is dotted by star symbol. Note that in the mirage point, the GUT scale 
stability cannot be satisfied.}
\label{stability}
\end{center}
\end{figure}

If $m_{3/2}$ is so large that the anomaly mediation contribution becomes dominant, the right-handed sleptons must have negative mass square\cite{anomaly}. 
Therefore, we have upper bound for the gravitino mass.
The upper bound for the ratio $m_{3/2}/M_{1/2}$ can be obtained by requiring that the positivity of the stop and stau mass squares at the SUSY breaking scale, or at the GUT scale.
From the eq. (\ref{eq:scalar_total}), the positivity condition for the right-handed stau mass square at $\mu$ can be written as
\begin{equation}
\ln\frac{\mu}{M_{\rm{mir}}}\leq \frac{10\pi}{33\alpha_1(\mu)}
\left[\sqrt{1+5.5\frac{\tilde m_{\tilde{\tau}_R}^2}{M_{1/2}^2}}-1\right].
\end{equation}
Since $\ln\frac{\mu}{M_{\rm{mir}}}=\ln\frac{\mu}{\Lambda}+\frac{m_{3/2}}{2M_{1/2}}$,
this gives the upper bound for $m_{3/2}$.
If we take $\tilde m_{\tilde{\tau}_R}=M_{1/2}$, the upper bound for the gravitino mass becomes $222 M_{1/2}$ for $\mu=1$ TeV and 76$M_{1/2}$ for $\mu=\Lambda_G$.
For the stop masses, numerical upper bounds are given in Fig. \ref{stability} for 
$\tilde A_t/M_{1/2}=-1, 1, 2$.
In the calculation, we assume that 
$\tilde m_{\tilde t_R}=\tilde m_{\tilde t_L}=\tilde m_{\tilde \tau_R}\equiv \tilde m$ and $\tilde m_{H_u}=0$.
All sfermion mass squares must be positive at least at the SUSY breaking scale. 
For this minimal requirement, roughly $m_{3/2} < 100 M_{1/2}$ if $\tilde m<M_{1/2}$, and $m_{3/2}< 500 M_{1/2}$ if $\tilde m<2M_{1/2}$.
If the positivity at the GUT scale is required (though this is not necessary for the consistency of the theory), $m_{3/2}< 200 M_{1/2}$ when $\tilde m<2M_{1/2}$.    
In numerical calculations in this paper, we take $m_t(\rm{pole})=173.07$ GeV and the unified gauge coupling $g_{\rm{GUT}}^2=0.48$.
We are interesting in the region $30<m_{3/2}/M_{1/2}<200$ in this paper.

Note that under the special boundary conditions $M_{1/2}=\tilde A_t=\sqrt{2}\tilde m_{\tilde{t}}$ in the mirage mediation the some sfermion mass squares become negative at the GUT scale as seen in the figure. However, in the general boundary conditions, the positivity at the GUT scale can be satisfied.

\subsection{Improvement in general cases}

In this subsection, we show that even in the general cases, the fine-tuning can be improved by using the numerical calculation.

First, we explain the improvement in the mirage mediation. 
Let us evaluate the quantum correction for the Higgs mass $m^2_{H_u}(\mu=1{\rm TeV})$ from the eq.(\ref{eq:mi_gravity}) obtained in the gravity mediation.
We express the quantum correction $\Delta m^2_{H_u}=m^2_{H_u}-\tilde{m}^2_{H_u}$ as
\begin{equation}
\Delta m^2_{H_u}({\rm 1 TeV})=c_0M_{1/2}^2+c_1\tilde{\Sigma}_t+c_2\tilde{A}_t^2+c_3\tilde{A}_tM_{1/2},
\label{eq:c}
\end{equation}
where constants $c_i$ are numerically calculated as
\begin{equation}
c_0=-1.601,\quad c_1=-0.396,\quad c_2=-0.082,\quad c_3=-0.260.
\end{equation}
If we set $M_{1/2}=\tilde{A}_t=\sqrt{\tilde{\Sigma}_t}$, we obtain $\Delta m^2_{H_u}=-2.34M_{1/2}^2$.
In order to obtain the quantum correction for the Higgs mass in the mirage mediation, we revaluate $c_i$ under the anomaly mediation from the eq. (\ref{eq:mi_anomaly}).
If we set $m_{3/2}/M_{1/2}=60.0$, we obtain 
\begin{equation}
c_0=0.291,\quad c_1=-0.396,\quad c_2=-0.082,\quad c_3=0.156.
\end{equation}
If we take the boundary conditions in the mirage mediation as 
$M_{1/2}=\tilde{A}_t=\sqrt{\tilde{\Sigma}_t}$, we obtain $\Delta m^2_{H_u}=-0.031M_{1/2}^2$.
These calculations show that the mirage mediation has more than one order less tuning is required than the gravity mediation without the anomaly mediation. 
The essential points for this improvement are that the coefficients $c_i$ become small and the cancellation happens because of the different signatures of $c_i$.

\begin{figure}[tbp]
\begin{center}
\includegraphics[scale=0.30]{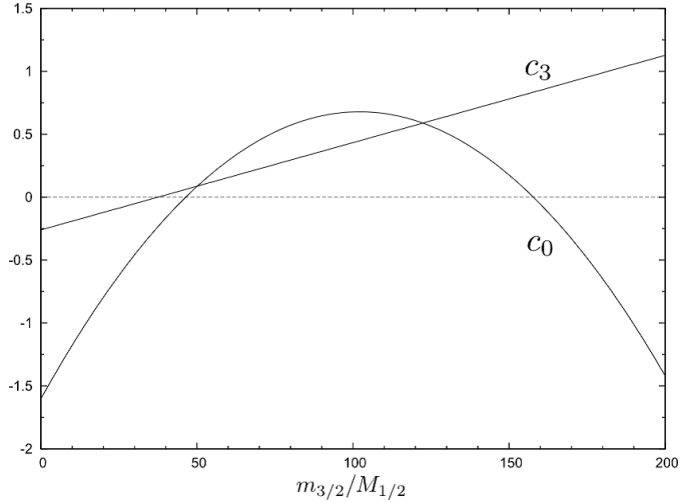}
\caption{Values of $c_0$ and $c_3$ in eq. (\ref{eq:c}) versus $m_{3/2}/M_{1/2}$.}
\label{figure:c}
\end{center}
\end{figure}

These points for the improvement are also applicable to the more general cases.
Therefore, it is obvious that even for the general cases, some improvement for the tuning can be expected at least when the ratio $m_{3/2}/M_{1/2}=60$.
Is this improvement realized only in this special value for the ratio?
Note that $c_1$ and $c_2$ do not change by including the anomaly mediation.
On the other hand, $c_0$ and $c_3$ depend on $M_{\mathrm{mir}}$, namely, $m_{2/3}/M_{1/2}$. 
Fig. \ref{figure:c} shows this dependence.
One can see that the absolute values of $c_0$ and $c_3$ are reduced by the anomaly mediation with wide range of value of $m_{2/3}/M_{1/2}$ among the range we are interested in.
Actually, if $29<m_{2/3}/M_{1/2}<73$, the condition $|c_i|<0.5$ is satisfied for $i=0,1,2,3$.
Therefore, we conclude that even in the general cases, some improvements for the tuning problem are expected in our scenario.

\begin{figure}[tbp]
\begin{center}
\includegraphics[scale=0.26]{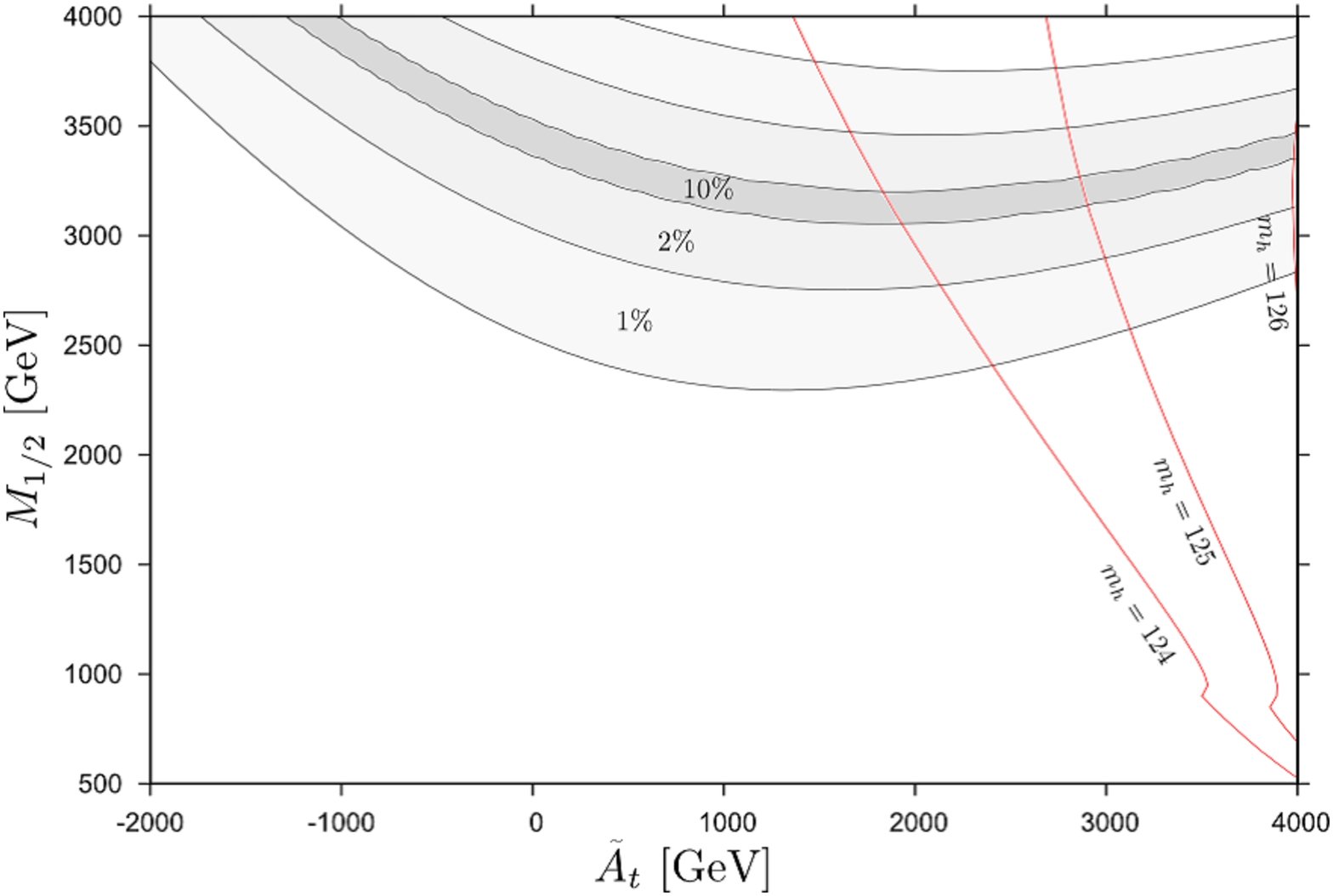}
\hspace{2mm}
\includegraphics[scale=0.26]{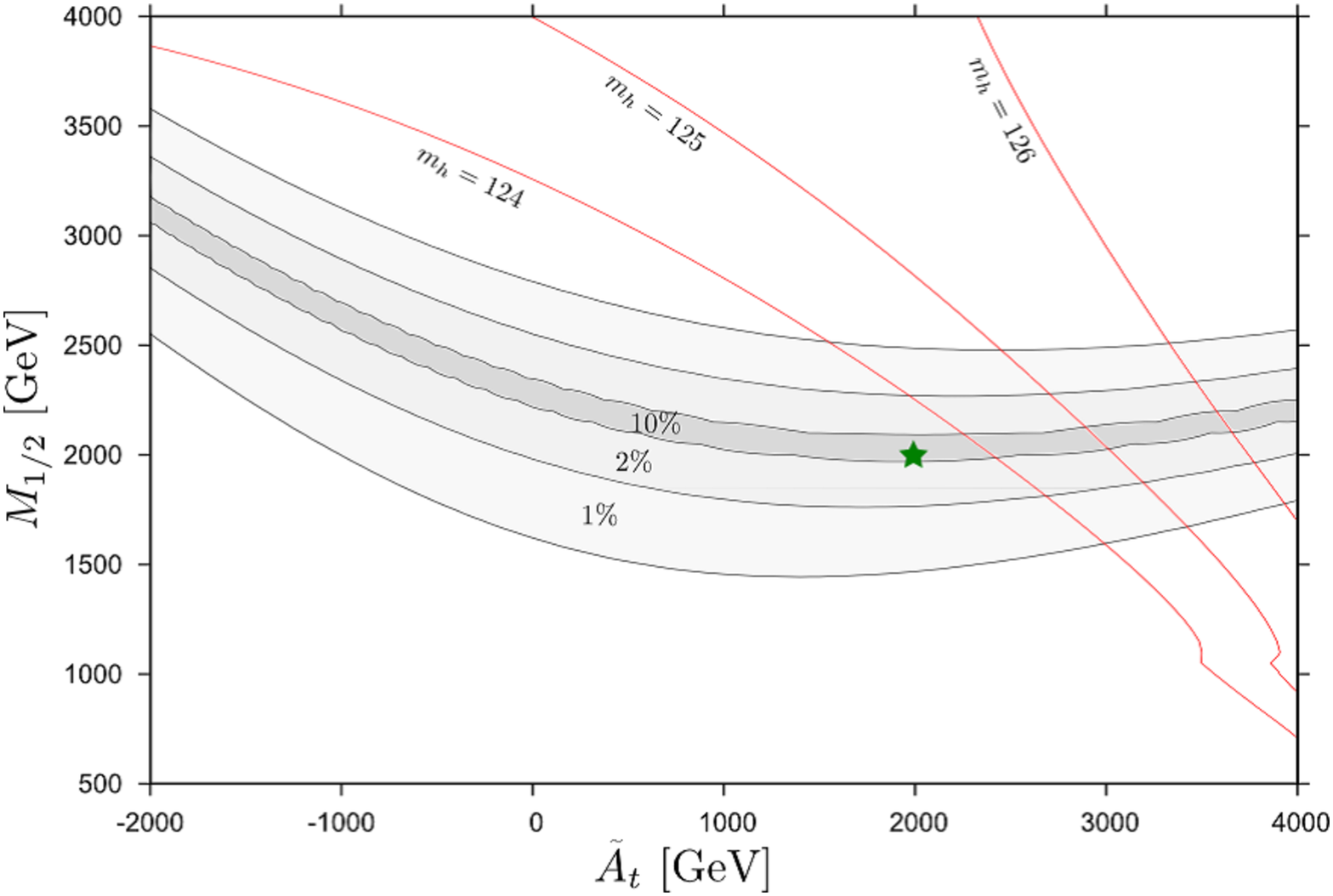}\\
\vspace{5mm}
\includegraphics[scale=0.26]{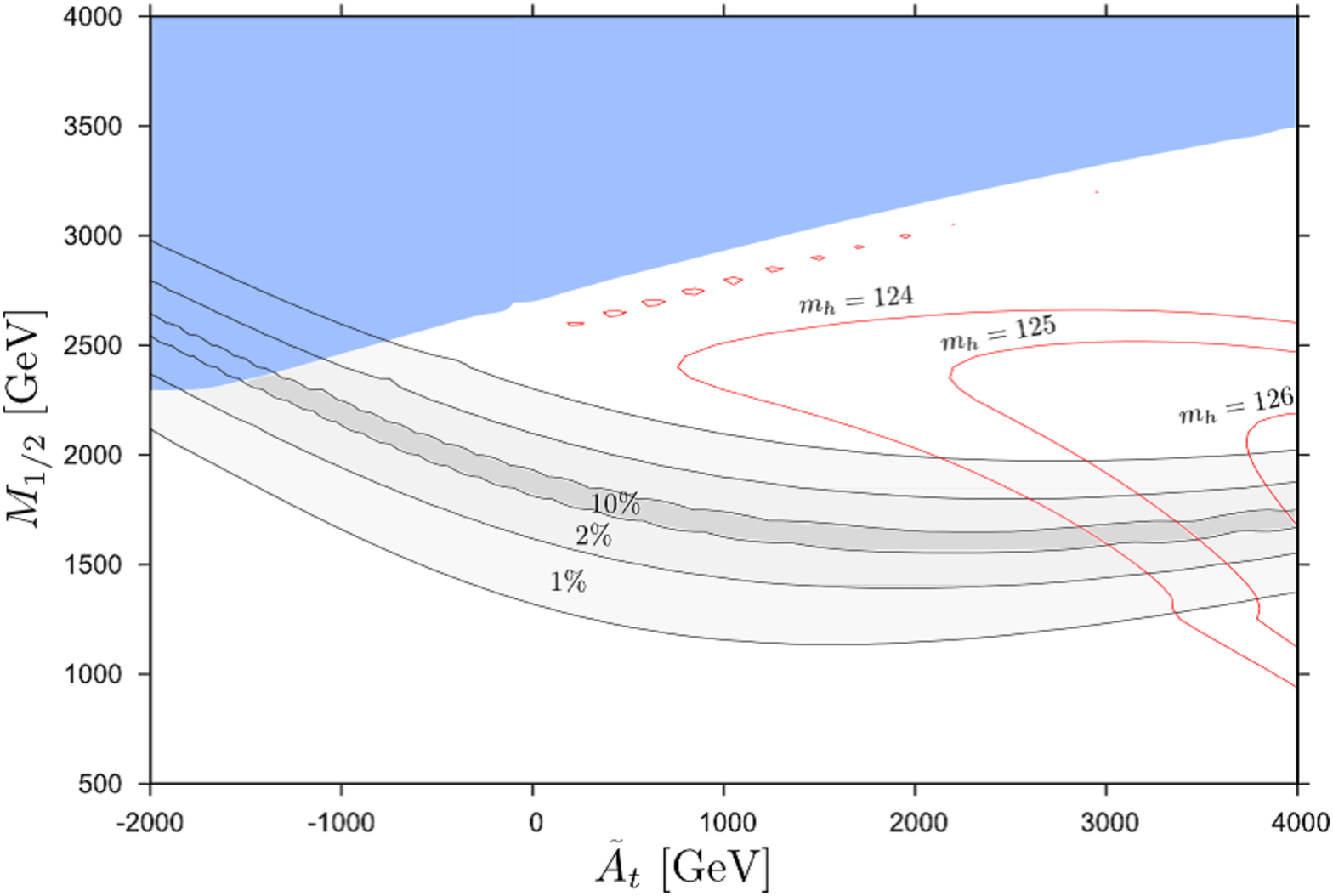}
\hspace{2mm}
\includegraphics[scale=0.26]{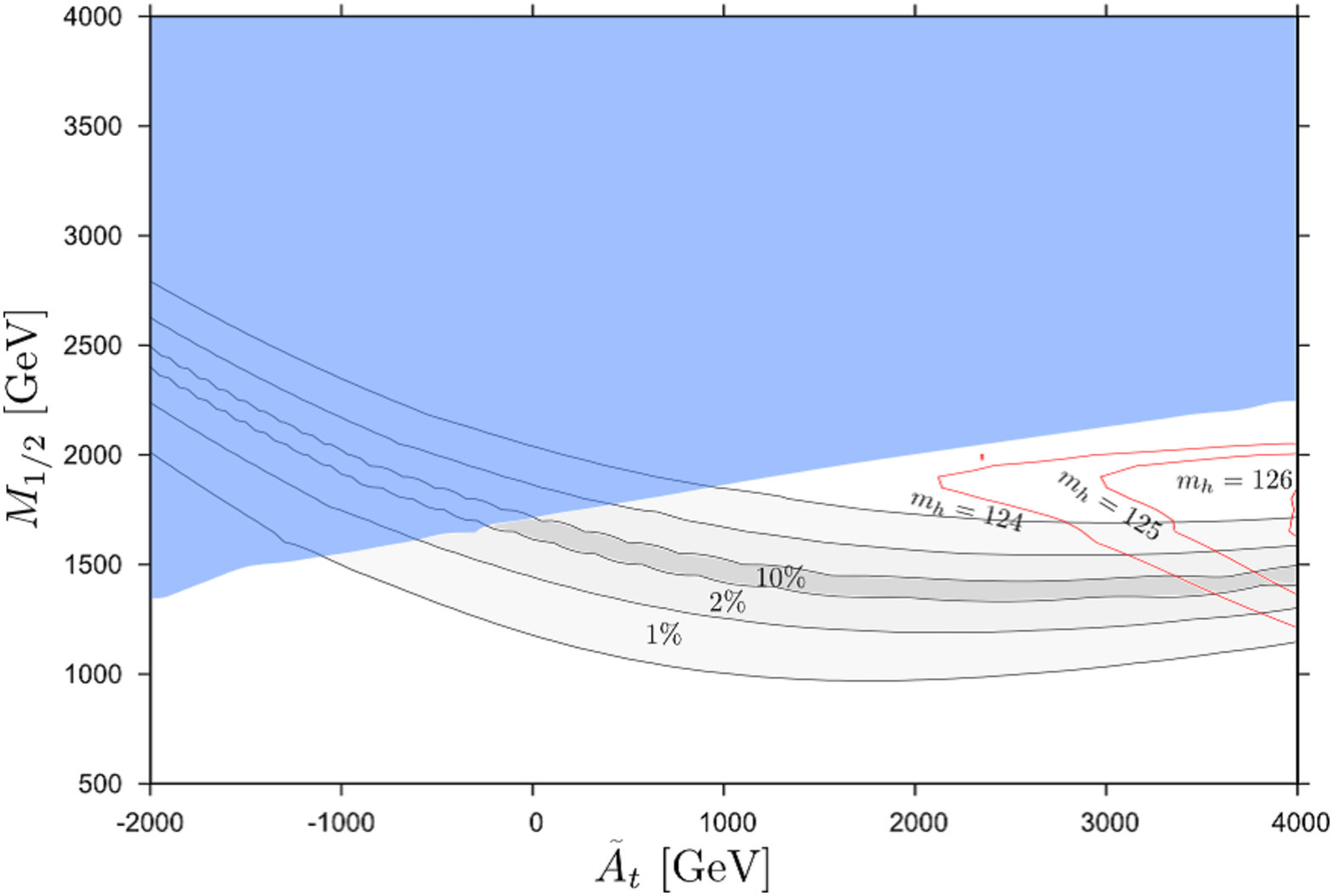}\\
\vspace{5mm}
\includegraphics[scale=0.26]{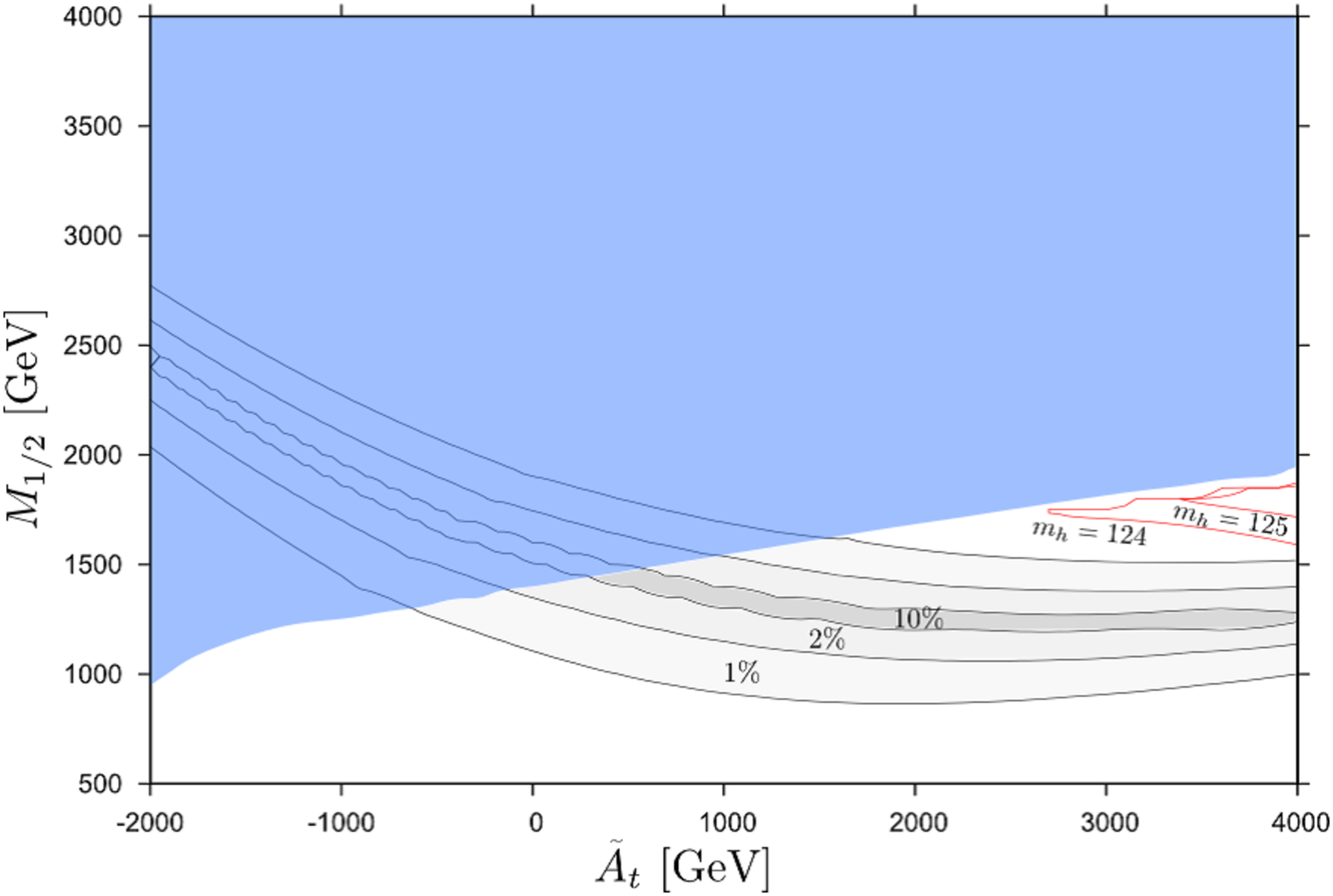}
\hspace{2mm}
\includegraphics[scale=0.26]{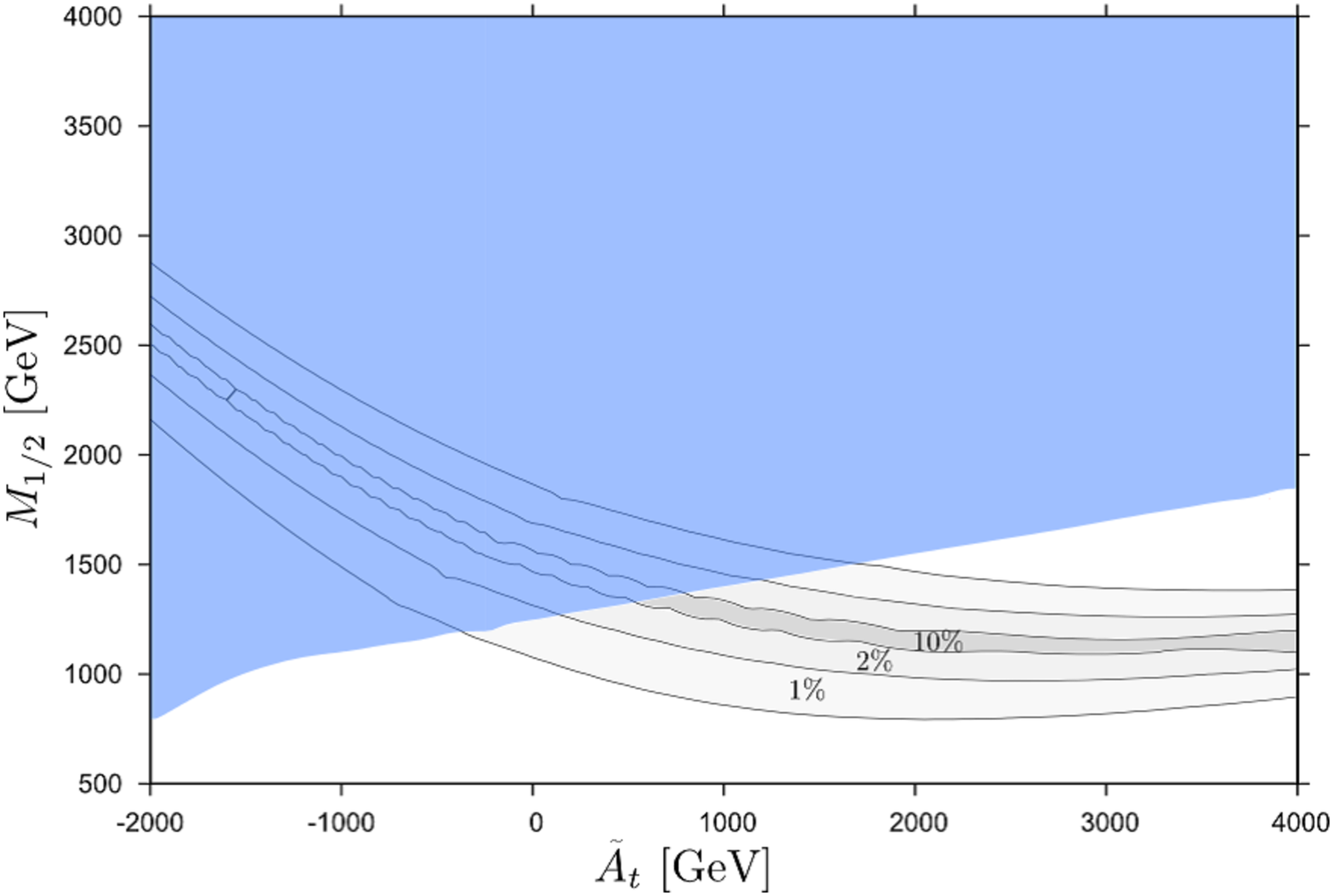}\\
\caption{$m_h^2/|2\Delta m_{Hu}^2(m_{\rm SUSY})|$ in 
$(\tilde A_t, M_{1/2})$ plain for $m_{3/2}/M_{1/2}=50$ (upper-left), 60 (upper-right), 70
(middle-left), 80(middle-right), 90(lower-left), 100(lower-right). 
We take $\sqrt{\tilde \Sigma_t}=2$ TeV. 
The shaded region is forbidden by the stability conditions at $m_{\rm{SUSY}}$.
For reference, the Higgs mass, which is calculated
by taking $\mu=m_A=500$ GeV, $\tan\beta=10$ and $m_0=3$ TeV, is shown as lines for 124 GeV, 125 GeV, and 126
 GeV. The mirage point is dotted by the star symbol. 
 If we require that $M_3(1{\rm TeV})>1$ TeV, $M_{1/2}$ must be larger than 813 GeV, 971 GeV, 
 1.22 TeV, 1.64 TeV, 2.44 TeV and 5.0 TeV for $m_{3/2}/M_{1/2}=50, 60, 70, 80, 90$ and 100, 
 respectively.
 The stability condition $m_{\tilde \tau_R}^2\geq 0$ at the GUT scale leads to the upper bound for $M_{1/2}$
 as 1.91 TeV, 1.69 TeV, 1.51 TeV, 1.38 TeV, 1.26 TeV, and 1.17 TeV for $m_{3/2}/M_{1/2}$=50, 
 60, 70, 80, 90, and 100, respectively. For large $\tilde A_t$, all sfermion mass squares can be
 positive till the GUT scale if this condition is satisfied.
}
\label{figure:2param}
\end{center}
\end{figure}

\begin{figure}[tbp]
\begin{center}
\includegraphics[scale=0.30]{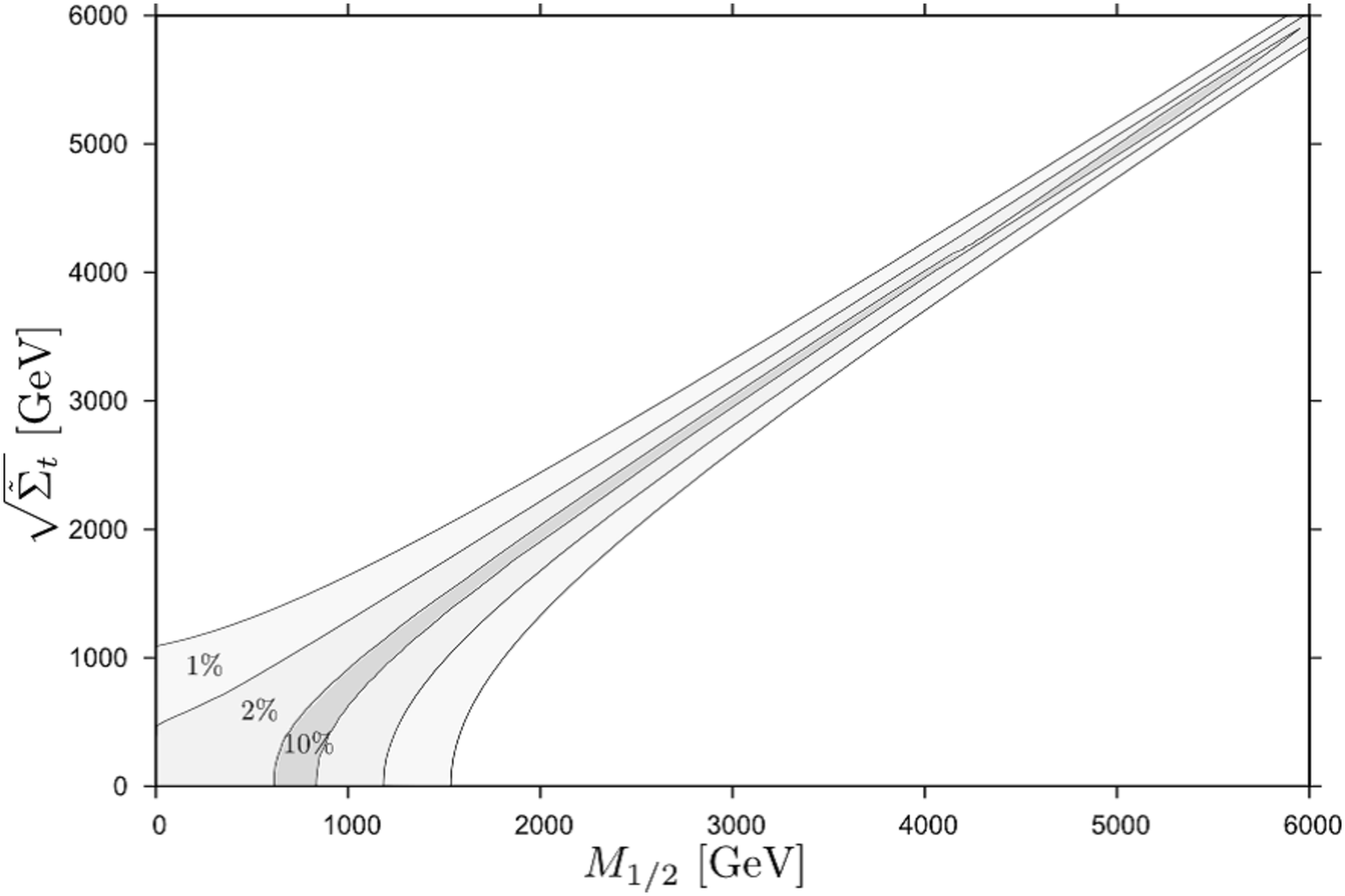}
\caption{$m_h^2/|2\Delta m_{Hu}^2(m_{\rm SUSY})|$ in 
$(M_{1/2}, \sqrt{\tilde\Sigma_t})$ plain. We take $\tilde A_t=2$ TeV and 
$m_{3/2}/M_{1/2}=60$. If we require $M_3(1\rm TeV)>1$ TeV, $M_{1/2}$ must be larger than 971 GeV.
}
\label{M1/2}
\end{center}
\end{figure}

Here we numerically check whether the quantum correction of the Higgs mass $\Delta m^2_{H_u}$ can be small.
In our scenario $\Delta m^2_{H_u}$ depends on four parameters: $M_{1/2}$, $\tilde{A}_t$, $\tilde{\Sigma}_t$ and $M_{\mathrm{mir}}$.
Hereafter we use $m_{3/2}/M_{1/2}$ instead of $M_{\mathrm{mir}}$.
Fig. \ref{figure:2param} shows $m_h^2/|2\Delta m^2_{H_u}(\mu=m_{\rm SUSY})|$ in 
$(\tilde A_t, M_{1/2})$ plain with $\sqrt{\tilde\Sigma_t}=2$ TeV which corresponds to
$\tilde m_{\tilde{t}_L}=\tilde m_{\tilde{t}_R}=\sqrt{2}$ TeV. Therefore, roughly, 
$m_{\rm SUSY}\sim \sqrt{2}$ TeV. 
The dark gray, gray and light gray regions are correspond to $(m_h^2/2)/|\Delta m^2_{H_u}|>$ 0.1, 0.02 and 0.01, respectively, where $m_h$ is the Higgs mass measured at the LHC as $m_h\sim 125$ GeV.
One can see that the tuning weaker than one percent is realized in a wide range of parameters. (Strictly speaking, we have to address how strong tuning is required for 
model parameters to be included in these areas. From these figures, we can 
see that $O(1\%)$ tuning is required in this scenario. Since this value is better
than $O(0.1\%)$ tuning for the usual minimal SUGRA boundary conditions, we can 
conclude that the tuning problem becomes less severe. )
Note that the amount of tuning for realizing small $\Delta m^2_{H_u}$ increases as $M_{1/2}$ becomes large as seen in Fig. \ref{M1/2}.
Therefore the masses of the gauginos should not be much larger than TeV scale if we expect not so large tuning with model parameters for getting small $\Delta m^2_{H_u}$.
One more important feature in general cases is that $A_t$ can be as large as $\sqrt{6}m_{\tilde t}$, which results in the maximal Higgs mass. This is an advantage in the general cases.
One can check from Fig. \ref{figure:2param} that $|\Delta m^2_{H_u}|$ is not influenced much by the value of $\tilde{A}_t$.
On the other hand, the large $A_t$ is important to obtain heavier Higgs. 
In the Fig. \ref{figure:2param}, we calculate the lightest Higgs mass by using the program
FeynHiggs-2.9.5 \cite{FeynHiggs}
under the additional assumptions which are adopted in the Ref. \cite{littlemirage2}.
Namely we asuume that $\tilde m_{\tilde{t}_L}^2=\tilde m_{\tilde{t}_R}^2=\tilde{\Sigma}_t/2$ and the parameters $\mu$, $\tan\beta$, $m_0$ and the mass of the CP odd Higgs $m_A$ are fixed by hand at the SUSY breaking scale.
(The latter assumption can be adopted if the unknown  GUT threshold corrections to the Higgs mass parameters are taken into account as noted in the Ref. \cite{littlemirage2}.)
Therefore we can realize the 125 GeV Higgs with small $m_{\tilde{t}}$ by setting $A_t/m_{\tilde{t}}\simeq \sqrt{6}$.
Actually when $50\leq m_{3/2}/M_{1/2}\leq90$, 125 GeV Higgs mass can be realized with reasonable value for $\tilde{A}_t$ as seen in Fig. \ref{figure:2param}.
On the other hand, if $m_{3/2}/M_{1/2}$ is 100, no line for 125 GeV Higgs appears because the stop masses are too small when $\tilde{m}_{\tilde{t}}=\sqrt{2}$ TeV. 
Note that these numerical results except for the Higgs mass can 
be basically obtained only from four parameters, $M_{1/2}$, $\tilde \Sigma_t$, 
$\tilde A_t$ and $m_{3/2}$. Once we fix the other parameters, we can discuss the other phenomenological
constraints from the LHC etc. 
Though it is important to show the allowed region for all parameters, it is beyond the scope
of our paper. Here, we just discuss the constraint from the gaugino masses which are determined by
$M_{1/2}$ and $m_{3/2}$ as
\begin{equation}
M_a(1{\rm TeV})\sim M_{1/2}\left[1+\frac{b_a\alpha_a}{2\pi}\left(-30+\frac{m_{3/2}}{2M_{1/2}}\right)\right]. 
\end{equation}
This is important since the gluino mass can be strongly constrained by the LHC experiments. 
We explicitly show $M_a/M_{1/2}$ at 1 TeV for various 
$m_{3/2}/M_{1/2}$ in Table \ref{gaugino}.
\begin{table}[tbp]
\begin{center}
\begin{tabular}{c|cccccccccccccc}
$\frac{m_{3/2}}{M_{1/2}}$ & 50  & 60 & 70 & 80 & 90 & 100 & 110 & 120 & 130 & 140 & 150 & 160  & 180  & 200 \\
\hline
$\frac{M_1}{M_{1/2}}$ & 0.90 & 0.99 & 1.08 & 1.17 & 1.26 & 1.35 & 1.44 & 1.53 & 1.62 & 1.71 & 1.80 & 1.89 &  2.07  & 2.25 \\
$\frac{M_2}{M_{1/2}}$ & 0.97 & 1.00 & 1.02 & 1.05 & 1.07 & 1.10 & 1.12 & 1.15 & 1.18 & 1.20 & 1.23 & 1.25 & 1.30  & 1,36 \\
$\frac{|M_3|}{M_{1/2}}$ &1.23&1.03& 0.82 & 0.61 & 0.41& 0.20 & 0.01 & 0.21 & 0.42 & 0.63 & 0.83 & 1.04 & 1.45  & 1.87 \\
\hline
\end{tabular}
\caption{$M_a/M_{1/2}$ for various $m_{3/2}/M_{1/2}$.  }
\label{gaugino}
\end{center}
\end{table}
Note that the gluino mass $M_3$ is vanishing around $m_{3/2}/M_{1/2}\sim 110$. 
This means that the LHC constraints from the gluino mass can be severe
if $m_{3/2}/M_{1/2}$ is around 110, though this cancellation is quite accidental.
Actually requiring $M_3>1$ TeV , we have no allowed region for $m_{3/2}/M_{1/2}=90$ and 
$\tilde m_{\tilde t}=\sqrt{2}$ TeV in Fig. \ref{figure:2param}. However, we should mension that
this result is strongly dependent on the value of $\tilde m_{\tilde t}$, because the stability condition
is essential. If we take larger  $\tilde m_{\tilde t}$, the larger $M_{1/2}$ is allowed and therefore
the allowed region must appear, though the finetuning must be severer.
But we have shown from this numerical calculation that we have sizable parameter region in which 
$O(1\%)$ tuning is realized. 
We do not have to take a special value for $m_{3/2}/M_{1/2}$ for obtaining  $O(1\%)$ tuning. 
This result is important.
 
Is it a general feature of this scenario that $\Delta m_{H_u}^2$ is dependent quite mildly on $A_t$?
This interesting feature can be understood also from the numerical formula 
(\ref{eq:c}), which is rewritten as
\begin{equation}
\Delta m^2_{H_u}({\rm 1 TeV})=\bar c_0M_{1/2}^2+
c_1\tilde{\Sigma}_t+c_2\left(\tilde{A}_t+\frac{c_3}{2c_2}M_{1/2}\right)^2,
\label{eq:cr}
\end{equation}
where $\bar c_0\equiv c_0-c_3^2/(4c_2)$. Since without the anomaly mediation contribution,
all parameters $\bar c_0$, $c_1$, and $c_2$ are negative,   
$\Delta m^2_{H_u}$ cannot be zero nor small and therefore tuning becomes
worse. However, if anomaly mediation contribution is sizable, 
$\bar c_0$ can be positive and therefore, $\Delta m^2_{H_u}$ can vanish. 
How large anomaly mediation contribution is needed for positive $\bar c_0$?
Numerically, $m_{3/2}\geq 47 M_{1/2}$ is needed.
What is important here is that $\Delta m_{H_u}^2$ is dependent on $\tilde A_t$ 
quite mildly when $\tilde A_t\sim -c_3M_{1/2}/(2c_2)$, which is derived from 
$\partial \Delta m^2_{H_u}/\partial \tilde A_t=0$.
The scale of $M_{1/2}$ for vanishing $\Delta m^2_{H_u}$ can be determind by cancellation condition for the first two terms in eq. (\ref{eq:cr}) as
$M_{1/2}\sim \sqrt{-c_1\tilde\Sigma_t/\bar c_0}=\sqrt{-2c_1/\bar c_0}\tilde m_{\tilde t}$. 
Note that the ratio $\tilde m_{\tilde t}/M_{1/2}=\sqrt{-\bar c_0/(2c_1)}$ is important in 
deriving the stability
conditions as in Fig. \ref{stability}. These values for various $m_{3/2}/M_{1/2}$ are found
in Table \ref{table:c}.
From both relations $\tilde A_t\sim -c_3M_{1/2}/(2c_2)$ and 
$M_{1/2}\sim \sqrt{-c_1\tilde\Sigma_t/\bar c_0}=\sqrt{-2c_1/\bar c_0}\tilde m_{\tilde t}$, 
an interesting relation $\tilde A_t/\tilde m_{\tilde t}=\sqrt{-c_1c_3^2/(2\bar c_0c_2^2)}$
is obtained. Surprisingly, in very wide range of $m_{3/2}/M_{1/2}$, the coefficient
$\sqrt{-c_1c_3^2/(2\bar c_0c_2^2)}$ is around 2 as in Table {\ref{table:c}}. 
This means that the interesting feature, that $\Delta m_{H_u}^2$ has quite mild dependence on 
$\tilde A_t$ around $\tilde A_t\sim 2$, and therefore we can obtain 125 GeV Higgs easier by 
taking large $\tilde A_t$, is generally realized in this scenario. 
\begin{table}[tbp]
\begin{center}
\begin{tabular}{c|ccccccccccc}
$\frac{m_{3/2}}{M_{1/2}}$ & 10 & 30 & 50  & 60 & 70 & 80 & 90 & 100 & 120 & 150 & 200 \\
\hline
$c_0$ & -1.177 & -0.458 & 0.085 & 0.291 & 0.453 & 0.572 & 0.646 & 0.677 & 0.608 & 0.176 & -1.418\\
$c_3$ & -0.191 & -0.052 & 0.087 & 0.156 & 0.225 & 0.295 & 0.364 & 0.433 & 0.572 & 0.780 & 1.127 \\
$\bar c_0$ &-1.066&-0.450& 0.108 & 0.365 & 0.607 & 0.837 & 1.050 & 1.249 & 1.606 & 2.031 & 2.454\\
$\sqrt{-\frac{\bar c_0}{2c_1}}$ &- &- & 0.369 & 0.679 & 0.876 & 1.023 & 1.152 & 1.256 & 1.425 & 1.603 & 1.761\\
$-\frac{c_3}{2c_2}$ &-1.165 &-0.317 &0.530 &0.951 &1.372 &1.799 &2.220 &2.640 &3.488 & 4.756 & 6.872\\
$\sqrt{-\frac{c_1c_3^2}{2\bar c_0c_2^2}}$ &- &- & 1.435 & 1.401 & 1.567 & 1.750 & 1.927 & 2.101 & 2.449 & 2.968 & 3.903\\
\hline
\end{tabular}
\caption{Coefficients $c_0$, $c_3$, etc. for various $m_{3/2}/M_{1/2}$. $\Delta m^2_{H_u}=0$ and
$\partial \Delta m^2_{H_u}/\partial \tilde A_t=0$ lead to 
$\sqrt{-\bar c_0/(2c_1)}=\tilde m_{\tilde t}/M_{1/2}$, $-c_3/(2c_2)=\tilde A_t/M_{1/2}$, 
respectively, and therefore $\sqrt{-c_1c_3^2/(2\bar c_0c_2^2)}=\tilde A_t/\tilde m_{\tilde t}$. }
\label{table:c}
\end{center}
\end{table}

\begin{figure}[tbp]
\begin{center}
\includegraphics[scale=0.30]{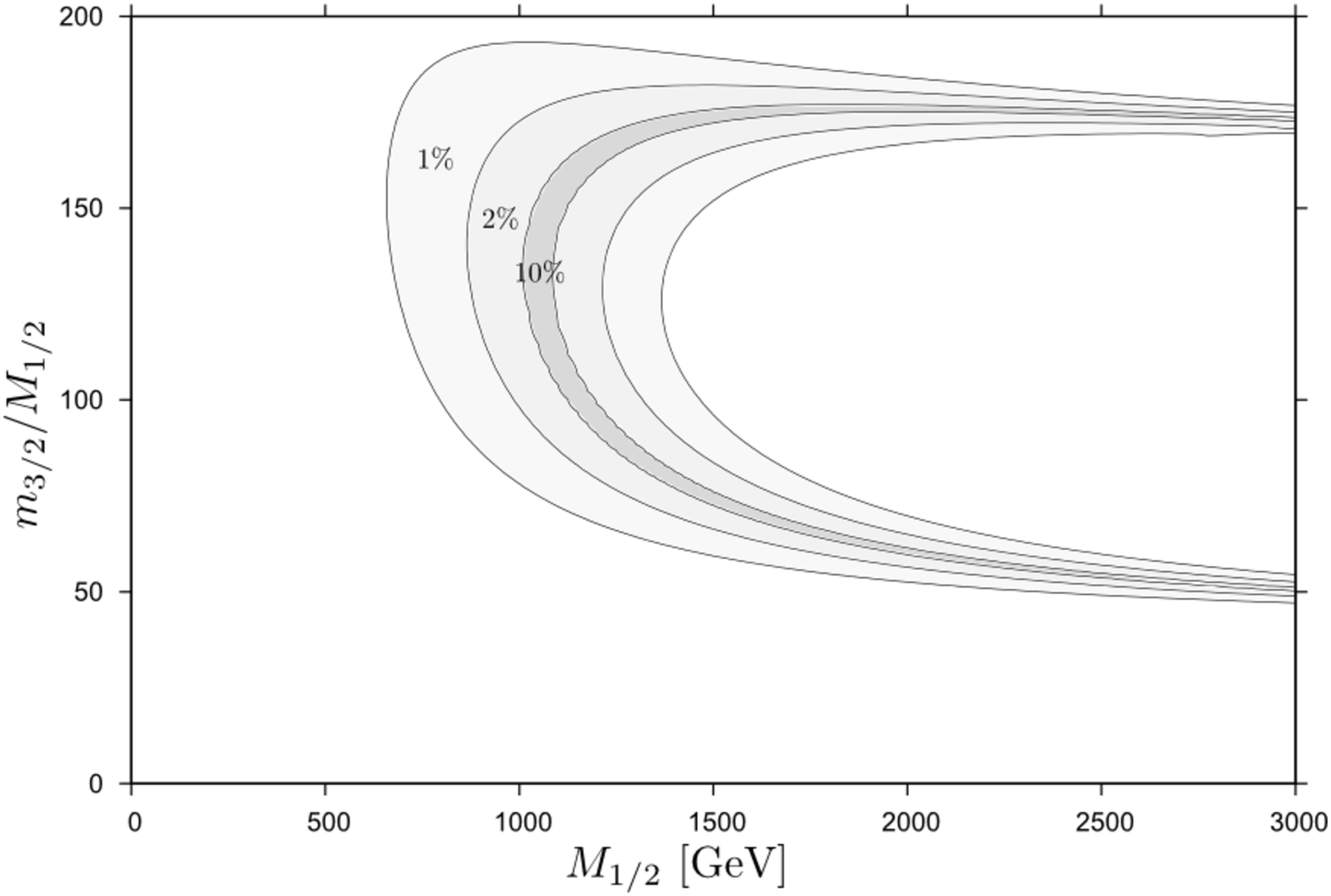}
\caption{$m_h^2/|2\Delta m_{Hu}^2(m_{\rm SUSY})|$ in 
$(M_{1/2}, m_{3/2}/M_{1/2})$ plain. We take $\tilde A_t=\sqrt{\tilde\Sigma_t}=2$ TeV.
}
\label{m32/M12}
\end{center}
\end{figure}

The lower bound for the ratio $m_{3/2}/M_{1/2}$ which realizes $\Delta m^2_{H_u}=0$ is also shown
in Fig. \ref{m32/M12} in which $\tilde A_t=\sqrt{\tilde\Sigma_t}=2$ TeV. This lower bound is
consistent with the above arguments from the numerical formula (\ref{eq:cr}).
Even the upper bound for the ratio $m_{3/2}/M_{1/2}$ is seen in Fig. \ref{m32/M12}.
The upper bound becomes lower than the value discussed in the above, because $\tilde A_t$ is
fixed to be 2 TeV in the numerical calculation in Fig. \ref{m32/M12}.
Interestingly the lower bound for $M_{1/2}$ is seen in the figure. 


\subsection{Strategy for testing GUT in general cases}

In this subsection, we discuss how to obtain the signatures for GUT scenarios from the mass spectrum of the SUSY particles, which is assumed to be observed by experiments here, in general cases.
As noted in the previous section, the top Yukawa contribution spoils the cancellation between the RG contribution and anomaly mediation contribution for the up-type Higgs mass and stop masses at $M_{\mathrm{mir}}$.
However, for the other sfermion masses and the gaugino masses, the cancellation at $M_{\mathrm{mir}}$ is still valid. 
Therefore, from the mass spectrum of the gauginos, we can obtain the mirage scale $M_{\mathrm{mir}}$ by calculating the RG equations for gaugino masses.
Once the mirage scale is known, we can obtain the gravity contribution to the masses of the sfermions other than two stops by calculating the RG equations from the SUSY breaking scale to the mirage scale.
It is beyond the scope of this paper how to test a concrete GUT scenario by this strategy.
We will study this subject in future.

\section{Summary and discussion}

We have shown that if we require that $m_{3/2}\sim O(100{\rm TeV})$ for solving the gravitino problem and the other SUSY breaking parameters are $O(1{\rm TeV})$ for the naturalness, the little hierarchy problem becomes less severe.
The essential point is that in such a situation, the anomaly mediation contribution becomes sizable, which can generically lower the messenger scale of the gravity mediation effectively.

If the Yukawa coupling is negligible, all the RG evolutions of gaugino mass, the scalar mass and trilinear coupling are canceled at the same scale $M_{\mathrm{mir}}$ by the anomaly mediation
contribution.
However, the Yukawa contribution breaks the complete cancellation at $M_{\mathrm{mir}}$ for the scalar and the trilinear coupling.
In practice the large top Yukawa coupling spoils the cancellation at $M_{\mathrm{mir}}$ for the stop masses, up-type Higgs mass, and $A_t$.
One possibility for vanishing the top Yukawa contribution is that the special boundary conditions are adopted such as the mirage mediation. 
This special boundary conditions are applied only for the stop masses and up-type Higgs masses, and therefore, we have no constraints for the other sfermion masses. 
First, we discussed the generalization of the mirage mediation. It is interesting that the natural SUSY mass spectrum is consistent with the mirage type boundary conditions.
Second, we have considered another possibility in which we do not have special boudary conditions for the gravity contributions. We have showed that even in such general cases, the tuning is improved in a wide range of paramter spaces we are interested in. 
An attractive feature of this scenario is that it has the flexibility of the mass parameters at the cutoff scale because we need not exactly cancel the top Yukawa contribution.
We can get large values like $A_t/m_{\tilde{t}}\simeq \sqrt{6}$, which is important for realizing 125 GeV Higgs with smaller $m_{\tilde{t}}$.

One of the disadvantage of the gravity mediation is that the universality of the sfermion masses, which are important in solving the SUSY FCNC problem, is not guaranteed generically.
One interesting possibility is to introduce flavor symmetry to realize the universality. 
One of the most interesting symmetries is $E_6\times SU(2)_F$ which realize the modified universality in which the third generation $\bf 10$ of $SU(5)$ can have different mass $m_3$ than the other sfermion mass $m_0$.
If we take $m_0\gg m_3\sim 1$ TeV, this is nothing but the natural SUSY type SUSY breaking parameters. 

Our new scenario has the several interesting features.
First, the mirage scale $M_{\mathrm{mir}}$, where the quantum corrections for the gaugino and the scalar masses which does not couple with top vanish, need not to be just the TeV scale.
The scale $M_{\mathrm{mir}}$ can be smaller than the weak scale, so long as the correction of the Higgs mass is not so large.
Then the lightest gaugino may be the gluino unlike the TeV-scale mirage mediation.

Second, this model predicts that the mass difference of two stop masses is around the weak scale, even if these masses are around the TeV scale.
Suppose two stop masses from the gravity mediation unifies at the cutoff scale
\begin{equation}
\tilde{m}^2_{\tilde{t}_L}=\tilde{m}^2_{\tilde{t}_R}.
\end{equation}
This is expected from the GUT models such as $SU(5)$.
The top Yukawa contribution splits these masses even at the mirage scale $M_{\mathrm{mir}}$.
However, these masses nearly degenerate if $\Delta m^2_{\tilde H_u}$ is small because the relation
\begin{equation}
\Delta m^2_{\tilde{t}_L}-\Delta m^2_{\tilde{t}_R}=-\frac{1}{3}\Delta m^2_{H_u}+\frac{M_{1/2}^2}{\pi}(-2\alpha_2+\frac{2}{5}\alpha_1)\ln\frac{\mu}{M_{\mathrm{mir}}}+\frac{M_{1/2}^2}{8\pi^2}(-4\alpha_2^2+\frac{132}{25}\alpha_1^2)\left(\ln\frac{\mu}{M_{\mathrm{mir}}}\right)^2
\end{equation}
can be found.
Note the QCD and top Yukawa contributions cancel between two stop masses, therefore the mass difference is approximately proportional to the correction of the Higgs mass.

In this paper, we have focused on the physics which can be discussed
by considering the specific parameters,
$M_{1/2}$, $\tilde\Sigma_t$
(or $m_3$), $\tilde A_t$ and $M_{\rm mir}$ (or $m_{3/2}$). 
Actually, all figures in this paper are based on these parameters except
in calculating the Higgs mass. 
However, in some cases, the other parameters can be important.
For example, it has been pointed out that when $m_0$ is much larger than $m_3$,
two loop RG effects give sizable negative contributions to stop mass square,
which makes the contraints in Fig. 1 more severe.
And of course, in order to disucuss phenomenological constraints from LHC, 
or FCNC processes, the other parameters must be fixed. 
For example, if we take $M_{1/2}=2$ TeV, $\tilde \Sigma_t=(2{\rm TeV})^2(\rightarrow \tilde m_{\tilde t}=\sqrt{2}$ TeV), $m_{3/2}/M_{1/2}=70$, 
$\tilde A_0=3.5$ TeV, $\tilde m_0=3$ TeV, $\tan\beta=10$ and $\mu=m_A=0.5$ TeV,
then, we can obtain the parameters at the scale $m_{\rm SUSY}=1130$ GeV as
$M_3=1630$ GeV, $M_2=2046$ GeV, $M_1=2162$ GeV,
$m_{\tilde {Q1}}=m_{\tilde {Q2}}=2743$ GeV,
$m_{\tilde{u_R}}=m_{\tilde{c_R2}}=2784$ GeV,
$m_{\tilde{d_R}}=m_{\tilde{s_R}}=m_{\tilde{b_R}}=2784$ GeV,
$m_{\tilde{L1}}=m_{\tilde{L2}}=m_{\tilde{L3}}=2948$ GeV,
$m_{\tilde{e_R}}=m_{\tilde{\mu_R}}=2979$ GeV,
$m_{\tilde{Q3}}=1012$ GeV,
$m_{\tilde {t_R}}=1252$ GeV,
$m_{\tilde {\tau_R}}=1370$ GeV,
$A_u=A_c=2981$ GeV,
$A_t=2095$ GeV,
$A_d=A_s=A_b=2693$ GeV,
$A_e=A_\mu=A_\tau=3317$ GeV and
$m_h=126.0$ GeV.
Phenomenological constraints from LHC can be satisfied in this example.
The constraint from the $b\rightarrow s\gamma$ may be sizable\cite{nagata}
but must
be milder because the stop and the chargino are heavier than in Ref. 
\cite{nagata}. (And the final allowed region is 
dependent on the SUSY mixing parameters which have not been fixed yet.) 
Though it is also important to show the allowed region with the all SUSY breaking parameters most of which have not been fixed in this paper,
this subject is beyond the scope of this paper. 

If the naturalness is required, the Higgsino mass $\mu$ must not be much larger than the weak scale.
Therefore, the lightest SUSY particle (LSP) can be expected to be the Higgsino. 
If it is additionally required that the thermally produced Higgsino abundance is consistent with the observed abundance of the dark matter, we can obtain further constraints on SUSY parameters.
We do not discuss this direction in detail.

One of the most important features in our scenario is that if the mirage scale is around the SUSY 
breaking scale, the signatures of the GUT scenarios can be observed directly by observing the mass spectrum of SUSY particles. It is difficult to reach the GUT scale directly by experiments 
while the SUSY GUT is the most promizing candidates as the physics beyond the SM. Therefore,
it becomes quite important that future experiments can observe the signature of the SUSY GUT, 
for example, through the $D$-term contributions to the sfermion masses.

\section*{Acknowledgments}

K.T. is supported by Grants-in-Aid for JSPS fellows.
N.M. is supported in part by Grants-in-Aid for Scientific Research from MEXT of 
Japan.


\begin{thebibliography}{99}
\bibitem{Aad:2012tfa}
  G.~Aad {\it et al.}  [ATLAS Collaboration],
  Phys.\ Lett.\ B {\bf 716} (2012) 1
  [arXiv:1207.7214 [hep-ex]].

\bibitem{Chatrchyan:2012ufa}
  S.~Chatrchyan {\it et al.}  [CMS Collaboration],
  Phys.\ Lett.\ B {\bf 716} (2012) 30
  [arXiv:1207.7235 [hep-ex]].

 \bibitem{gravitino}
  S.~Weinberg,
  Phys.\ Rev.\ Lett.\  {\bf 48}, 1303 (1982);
  I.~V.~Falomkin, G.~B.~Pontecorvo, M.~G.~Sapozhnikov, M.~Y.~Khlopov, F.~Balestra and G.~Piragino,
  Nuovo Cim.\  A {\bf 79} (1984) 193
  [Yad.\ Fiz.\  {\bf 39} (1984) 990];
  M.~Y.~Khlopov and A.~D.~Linde,
  Phys.\ Lett.\  B {\bf 138} (1984) 265;
  J.~R.~Ellis, J.~E.~Kim and D.~V.~Nanopoulos,
  Phys.\ Lett.\  B {\bf 145}, 181 (1984);
  M.~Kawasaki and T.~Moroi, Prog. Theor. Phys. {\bf 93}, 879 (1995).

\bibitem{gravitino2}
  M.~H.~Reno and D.~Seckel, Phys.~Rev. D {\bf 37}, 3441 (1988); 
  K.~Kohri, Phys.~Rev. D {\bf 64}, 043515 (2001) [astro-ph/0103411];
  M.~Kawasaki, K.~Kohri and T.~Moroi, Phys.~Lett. B {\bf 625}, 7 (2005) [astro-ph/0402490];
  M.~Kawasaki, K.~Kohri, T.~Moroi and A.~Yotsuyanagi, Phys. Rev. D {\bf 78}, 065011 (2008) 
  [arXiv:0804.3745].
    
\bibitem{HighScale}
 J.~D.~Wells, [arXiv:hep-ph/0306127];
 N.~Arkani-Hamed and S.~Dimopoulos, JHEP {\bf 0506}, 073 (2005) [hep-th/0405159];
 G.~F.~Giudice and A.~Romanino, Nucl.\ Phys. B {\bf 699}, 65 (2004) [Erratum-ibid. B {\bf 706}, 65
 (2005) [hep-ph/0409232]:
 J.~D.~Wells, Phys.\ Rev. D {\bf 71}, 015013 (2005) [hep-ph/0411041].

\bibitem{HighScale2}
 G.~F.~Giudice and A.~Strumia, Nucl.~Phys. B {\bf 858}, 63 (2012) [arXiv:1108.6077];
 L.~J.~Hall and Y.~Nomura, JHEP {\bf 1201}, 082 (2012) [arXiv:1111.4519];
 L.~J.~Hall, Y.~Nomura and S.~Shirai, JHEP {\bf 1301}, 036 (2013) [arXiv:1210.2395];
 M.~Ibe and T.~T.~Yanagida, Phys.\ Lett. B {\bf 709}, 374 (2012) [arXiv:1112.2462];
 M.~Ibe, S.~Matsumoto and T.~T.~Yanagida, Phys.\ Rev. D {\bf 85}, 095011 (2012) [arXiv:1202.2253];
 A.~Arvanitaki, N.~Craig, S.~Dimopoulos and G.~Villadoro, JHEP {\bf 1302}, 126 (2013) [arXiv:1210.0555].

\bibitem{mirage}
 K.~Choi, A.~Falkowski, H.~P.~Nilles, M.~Olechowski and S.~Pokorski, JHEP {\bf 0411}, 076 (2004) 
 [hep-th/0411066]; 
 K.~Choi, A.~Falkowski, H.~P.~Nilles and M.~Olechowski, Nucl. Phys. B {\bf 718} (2005) [hep-th/0503216];
 K.~Choi, K.~S.~Jeong and K.~i.~Okumura, JHEP {\bf 0509}, 039 (2005) [hep-ph/0504037];
 M.~Endo, M.~Yamaguchi and K.~Yoshioka, Phys. Rev. D {\bf 72}, 015004 (2005) [hep-ph/0504036].

 \bibitem{mirage2}
 K.~Choi, K.~S.~Jeong, S.~Nakamura, K.-I.~Okumura, and M.~Yamaguchi, JHEP {\bf 0904}, 107 (2009) 
 [arXiv:0901.0052].
\bibitem{moduli}
 V.~S.~Kaplunovsky and J.~Louis, Phys. Lett. B {\bf 306}, 269 (1993) [hep-th/9303040];
 A.~Brignole, L.~E.~Ibanez and C.~Munoz, Nucl. Phys. B {\bf 422}, 125 (1994) [Erratum-ibid. B {\bf 436},
 747 (1995)][hep-ph/9308271];
 T.~Kobayashi, D.~Suematsu, K.~Yamada and Y.~Yamagishi, Phys. Lett. B {\bf 348}, 402 (1995)
  [hep-ph/9408322];
 L.~E.~Ibanez, C.~Munoz and S.~Rigolin, Nucl. Phys. B {\bf 553}, 43 (1999) [hep-ph/9812397].

\bibitem{anomaly}
 L.~Randall and R.~Sundrum, Nucl. Phys. B {\bf 557}, 79 (1999) [hep-th/9810155];
 G.~F.~Giudice, M.~A.~Luty, H.~Murayama and R.~Rattazzi, JHEP {\bf 9812}, 027 (1998) [hep-ph/9810442].

\bibitem{littlemirage}
 K.~Choi, K.~S.~Jeong, T.~Kobayashi and K.~i.~Okumura, Phys. Lett. B {\bf 633}, 355 (2006) 
 [hep-ph/0508029];
  K.~Choi, K.~S.~Jeong, T.~Kobayashi and K.~i.~Okumura, Phys. Rev. D {\bf 75}, 095012 (2007)
 [hep-ph/0612258].
\bibitem{littlemirage2}
 R.~Kitano and Y.~Nomura, Phys. Lett. B {\bf 631}, 58 (2005) [hep-ph/0509039].

\bibitem{NMSSM}
 P.~Fayet, Nucl. Phys. B {\bf 90}, 104 (1975); Phys. Lett. B {\bf 64}, 159 (1976); Phys. Lett. B {\bf 69},
 489 (1977); Phys. Lett. B {\bf 84}, 416 (1979); H.~P.~Nilles, M.~Srednicki and D.~Wyler, Phys. Lett. B
 {\bf 120}, 346 (1983); J.~M.~Frere, D.~R.~T.~Jones and S.~Raby, Nucl. Phys. B {\bf 222}, 11 (1983);
  J.~P.~Derendinger and C.~A.~Savoy, Nucl. Phys. B {\bf 237}, 307 (1984);
 J.~R.~Ellis, J.~F.~Gunion, H.~E.~Haber, L.~Roszkowski and F.~Zwirner, Phys. Rev. D {\bf 39}, 844 (1989);
 M.~Drees, Int. J. Mod. Phys. A {\bf 4}, 3635 (1989).

\bibitem{gauge}
 M.~Dine, W.~Fischler and M.~Srednicki, Nucl. Phys. B {\bf 189}, 575 (1981);
 S.~Dimopoulos and S.~Raby, Nucl. Phys. B {\bf 192}, 353 (1981);
 M.~Dine and W.~Fischler, Phys. Lett. B {\bf 110}, 227 (1982);Nucl. Phys. B {\bf 204}, 346 (1982);
 C.~R.~Nappi and B.~A.~Ovrut, Phys. Lett. B {\bf 113}, 175 (1982);
 L.~Alvare-Gaume, M.~Claudson and M.~B.~Wise, Nucl. Phys. B {\bf 207}, 96 (1982):
 S.~Dimopoulos and S.~Raby, Nucl. Phys. B {\bf 219}, 479 (1983);
 I.~Affleck, M.~Dine and N.~Seiberg, Nucl. Phys. B {\bf 256}, 557 (1985);
 M.~Dine and A.~E.~Nelson, Phys. Rev. D {\bf 48}, 1277 (1993) [hep-ph/9303230];
 M.~Dine, A.~E.~Nelson and Y.~Shirman, Phys. Rev. D {\bf 51}, 1362 (1995) [hep-ph/9507378];
 M.~Dine, A.~E.~Nelson, Y.~Nir and Y.~Shirman, Phys. Rev. D {\bf 53}, 2658 (1996) [hep-ph/9507378].
 
\bibitem{maximal}
M.~Carena, S.~Heinemeyer, C.~Wagner and G.~Weiglein, Eur. Phys. J. C{\bf 26}, 601 (2003) [hep-ph/0202167].

\bibitem{NaturalSUSY}
A.~Cohen, D.~Kaplan and A.~Nelson, Phys. Lett. B {\bf 388}, 588 (1996) [hep-ph/9607394];
N.~Arkani-Hamed and H.~Murayama, Phys. Rev. D {\bf 56}, 6733 (1997) [hep-ph/9703259].


\bibitem{Naturalmirage}
S.~Krippendorf, H.~P.~Nilles, M.~Ratz, M.~W.~Winkler, Phys. Lett. B {\bf 712}, 87 (2012)
 [arXiv:1201.4857];
M.~Asano and T.~Higaki, Phys. Rev. D {\bf 86}, 035020 (2012) [arXiv:1204.0508].

\bibitem{E6SU2}
N.~Maekawa, Phys. Lett. B {\bf 561}, 273 (2003) [hep-ph/0212141]; Prog, Theor. Phys. {\bf 112}, 639 (2004)
[hep-ph/0402224]; 
M.~Ishiduki, S.~-G.~Kim, N.~Maekawa, and K.~Sakurai, Prog. Theor. Phys. {\bf 122}, 659 (2009) 
[arXiv:0901.3400]; Phys. Rev. D {\bf 80}, 115011 (2009) [arXiv:0910.1336];
H.~Kawase and N.~Maekawa, Prog. Theor. Phys. {\bf 123}, 941 (2010) [arXiv:1005.1049];
N.~Maekawa and K.~Takayama, Phys. Rev. D {\bf 85}, 095015 (2012) [arXiv:1202.5816];
N.~Maekawa and Y.~Muramatsu, [arXiv:1401.2633].

\bibitem{KMY}
Y.~Kawamura, H.~Murayama, and M.~Yamaguchi, Phys. Rev. D {\bf 51}, 1337 (1995) [hep-ph/9406245].

\bibitem{FeynHiggs}
  S.~Heinemeyer, W.~Hollik and G.~Weiglein,
  Comput.\ Phys.\ Commun.\  {\bf 124}, 76 (2000)
  [hep-ph/9812320];
  Eur.\ Phys.\ J.\ C {\bf 9}, 343 (1999)
  [hep-ph/9812472];
  G.~Degrassi, S.~Heinemeyer, W.~Hollik, P.~Slavich and G.~Weiglein,
  Eur.\ Phys.\ J.\ C {\bf 28}, 133 (2003)
  [hep-ph/0212020];
  M.~Frank, T.~Hahn, S.~Heinemeyer, W.~Hollik, H.~Rzehak and G.~Weiglein,
  JHEP {\bf 0702}, 047 (2007)
  [hep-ph/0611326].

\bibitem{nagata}
  K.~Ishiwata, N.~Nagata and N.~Yokozaki,
  Phys.\ Lett.\ B {\bf 710} (2012) 145
  [arXiv:1112.1944].


\end{thebibliography}
\end{document}